\documentclass[12pt]{article}
\usepackage{amssymb}
\usepackage{graphicx}
\usepackage{epsf}
\usepackage{lscape}
\usepackage{epstopdf}
\usepackage[cp866]{inputenc}
\usepackage[T2A]{fontenc}
\usepackage[english]{babel}
\usepackage{amsmath}
\begin{document}

\title{\large{\bf   Asymptotic theory of charged particle transfer reactions at low energies   and    nuclear astrophysics   }}
\author{    R. Yarmukhamedov,$^{\rm{1,\,4},}$\thanks{Corresponding author, E-mail:
rakhim@inp.uz} \,K. I. Tursunmakhatov,$^{\rm{2}}$ and N. Burtebayev$^{\rm{3,\,4}}$ }

\maketitle{\it
$^{\rm{1}}$Institute of Nuclear Physics, Uzbekistan Academy of Sciences,\\ Tashkent 100214, Uzbekistan\\
$^{\rm{2}}$ Physical  and Mathematical Department of Gulistan State University, \\ 120100 Gulistan city, Uzbekistan\\
$^{\rm{3}}$Institute of Nuclear Physics, 050032 Almaty,  Kazakhstan\\
$^{\rm{4}}$  Al Farabi Kazakh National University,\\ Almaty, 050040, Kazakhstan\\ }
\bigskip

\begin{abstract}

 A new asymptotic theory is proposed for the peripheral sub- and above-barrier transfer $A$($x$, $y$)$B$  reaction within the three-body ($A$, $a$ and $y$) model, where $ x$= $y$ + $a$, $B$= $A$ + $a$ and $ a$  is a transferred particle.  In the asymptotic  theory,  the  allowance  of  the contribution of the three-body ($A$, $a$ and $y$) Coulomb  dynamics of the  transfer mechanism  to the peripheral partial amplitudes for the partial wave $l_i>>$ 1 and of  the Coulomb-nuclear distorted effects in the entrance and exit channels is done in a  correct manner within the framework of the dispersion theory and   the conventional distorted-wave Born approximation (DWBA), respectively.  It is shown that the proposed asymptotic theory  makes it possible    to test  the accuracy of taking into account  the the three-body Coulomb  dynamics of  the transfer mechanism in  the modified  DWBA.        The results of the  analysis of  the   differential cross sections of the specific proton and  triton  transfer reactions at  above- and sub-barrier energies are
 presented. New estimates and their uncertainties are obtained for values of the  asymptotic normalization coefficients for ${\rm{^9Be}} \,+p \to\,{\rm{^{10}B}}$, ${\rm{^{11}B}}\,+\,p\to\,{\rm{^{12}C}}$, ${\rm{^{16}O}} \,+\,p \to\, {\rm{^{17}F}}$ and ${\rm{^{19}F}}\to\, {\rm{^{16}O}} \,+\,t$ as well as for  the  direct  astrophysical  $S$ factors at stellar  energy of the   radiative capture   ${\rm{^9Be(}}p,\,\gamma{\rm{)^{10}B}} $,  ${\rm{^{11}B(}}p,\,\gamma{\rm{)^{12}C}} $ and  ${\rm{^{16}O(}}p,\,\gamma{\rm{)^{17}F}}$ reactions.  
\end{abstract}

\newpage

 \vspace{1.0cm}
\begin{center}  
 {\bf I. INTRODUCTION}
 \end{center}
  
\bigskip

In the  last two decades,  a number of    methods of  analysis of  experimental data for different nuclear processes   were proposed  to obtain  information on  the ``indirect determined'' (``experimental'') values of  the specific asymptotic normalization coefficients (or respective nuclear vertex constants) with the aim of their application to   nuclear astrophysics (see, for example, Refs. [1--5] and the available references therein).     One of such methods uses the modified DWBA \cite{Ar96,Mukh2}   for nuclear transfer reactions of manifest peripheral character in which   the differential cross sections are    expressed in the terms of the asymptotic normalization coefficients. One notes that an   asymptotic normalization coefficient  (ANC), which is proportional to the nuclear vertex constant (NVC) for the virtual decay $B\,\to\,A\,+\,a$,  determines the amplitude of the tail of the overlap function corresponding  to the wave function of nucleus $B$ in the binary  ($A\,+\,a$) channel (denoted by $A\,+a\,\to\, B$ everywhere below) \cite{Blok77}.  As   the ANC for $ A\,+a\,\to\, B$ determines the probability of the configuration $A+a$ in nucleus $B$ at distances greater than the radius of nuclear $Aa$ interaction,     the ANC arises   naturally in  expressions for the cross sections of  the peripheral   nuclear reactions between charged particles at low energies, in particular, of the peripheral exchange    $A{\rm{(}}B,\,A{\rm{)}}B$, transfer $A$($x$, $y$)$B$ and  astrophysical nuclear $A(a,\,\gamma)B$ reactions.    

\begin{figure}
\begin{center}
\includegraphics[height=4cm, width=14cm]{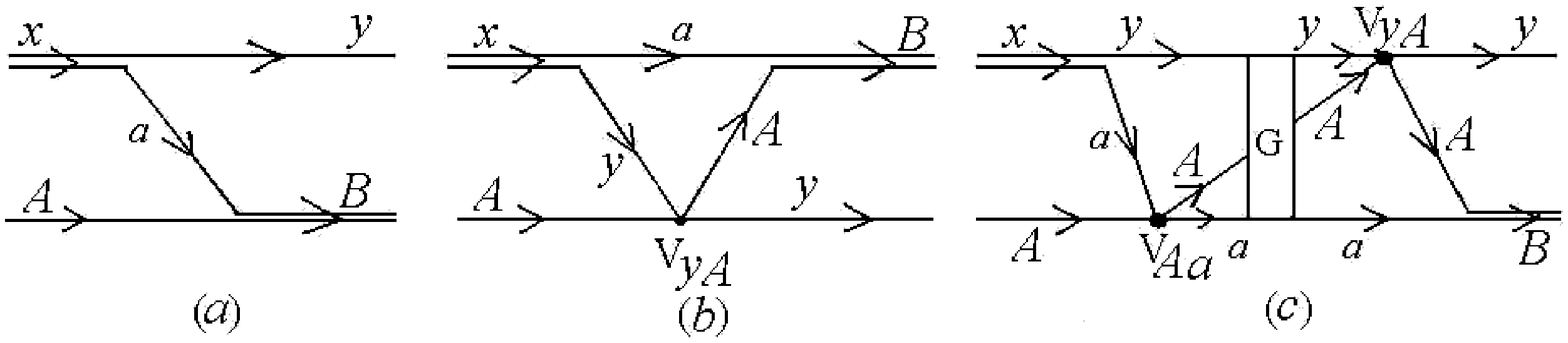}
 \end{center}
\caption{\label{fig1} Diagrams describing transfer of the particle $a$ and taking into account possible subsequent Coulomb-nuclear rescattering of particles ($A$, $a$ and $y$) in the  intermediate state.}
\end{figure}

 In the present work, the peripheral charged particle  transfer reaction
\begin{equation}
\label{subeq1}  
x\,+\,A\,\longrightarrow \,y\,+\,B
\end{equation}
is considered in the framework  of the three-body ($A$, $a$ and $y$) model,   where $x$=($y\,+\,a$) is a projectile,  $B$=($A\,+\,a$) and  $a$ is a transferred particle.  The main  idea       is based on   the following two  assumptions: i) the   peripheral  reaction (\ref{subeq1}) is  governed by the singularity  of the  reaction  amplitude   at $\cos\theta=\,\xi>$ 1, where $\xi$ is the  nearest to physical (-1 $\le\cos\theta\le$ 1) region  singularity generated by the pole mechanism 
(Fig. \ref{fig1}$a$) \cite{Shapiro}   and $\theta$ is the scattering angle in the center of mass; ii)  the dominant pole played by this nearest  singularity  is the result of the peripheral nature of the considered reaction   at least in the main peak of the angular distribution \cite{DDM1973}. Consequently,  it is necessary to know  the  behavior of the  reaction  amplitude at the  nearest singularity $\xi$ \cite{Av86,Mukh10}, which in turn  defines the behavior of the true  peripheral partial amplitudes   at  $l_i\,\gtrsim L_0>>$ 1 
($L_0\sim k_iR^{{\rm{ch}}}_i$ with  $R_i^{{\rm{ch}}}\,\gtrsim\, R_N$) \cite{Popov1964}  giving the dominant contribution to the reaction amplitude  at least in the main peak of the angular distribution \cite{DDM1973,KMYa1988}, where   $l_i$, $k_i$,  $R^{ch}_i$ and  $R_N$   are a  partial  wave,  a  number wave (or a relative momentum),  a channel radius,  and  the radius of the nuclear interaction       of  the colliding nuclei, respectively.

 In practice,  the ``post''-approximation  and  the ``post'' form   of  the modified DWBA   \cite{Ar96,Mukh2}    are used for the analysis  of the specific peripheral proton transfer reactions. They are restricted  by the zero- and first-order terms of the  perturbation theory   over  the optical Coulomb polarization operator  $\Delta V_f^C$ (or $\Delta V_i^C$) in   the transition operator,  respectively, which   are   sandwiched by the initial and final state wave functions  in  the matrix element of the reaction (\ref{subeq1}). At this, it is assumed  that the contribution of the first-order term over $\Delta V_f^C$ (or $\Delta V_i^C$)
 to   the matrix element is  small \cite{Mukh2}. But it was shown in Refs. \cite{Yarm2013,Mukh10,Yar97,Igam072}  that, when the residual nuclei  $B$ are formed in
weakly bound states being  astrophysical interest, this assumption is not guaranteed for the peripheral
charged particle  transfer reactions and, so,
the extracted ``experimental''  ANC values   may not have
 the necessary accuracy for their  astrophysical application (see, for example, \cite{Igam072} and Table 1 in  \cite{Yarm2013}). In this case, in
the transition operator an inclusion of all other   orders (the  second and higher orders) of the power expansion in a series over $\Delta V_f^C$ (or $\Delta V_i^C$) is required  for the DWBA cross section calculations since they  strongly change the power of the peripheral partial amplitudes at $l_i\,>>$ 1 \cite{Mukh10,Igam072}.    

For these reasons,    it is of great   interest to derive the expressions  for the amplitude and the  differential cross section (DCS) of the peripheral  reaction (\ref{subeq1}) within the so-called hybrid theory: the  DWBA approach and the dispersion peripheral   model \cite{DDM1973,Av86}. The main advantage of the hybrid theory  as compared to the modified DWBA  is that, first,  it   allows one  to derive  the expression for the part of  the reaction amplitude having  only the contribution from  the nearest singularity $\xi$ in which  the influence  of   the three-body Coulomb dynamics of the  transfer mechanism on the peripheral partial amplitudes at $l_i>>$ 1    is taken into account in a correct manner within the dispersion theory. Second, it accounts for the distorted effects in the initial and final states within  the DWBA approach, which is  more accurate than  as it was done     in \cite{DMYa1978} in   the dispersion peripheral model \cite{DDM1973}. They allow one to treat the important issue: to what extent does a  correct taking into account of   the  three-body Coulomb effects in the initial, intermediate and final states  of the peripheral reaction (\ref{subeq1}), firstly,  influence  the spectroscopic information deduced from the analysis of the experimental DCSs   and, secondly,   improve  the accuracy of the modified DWBA analysis  used for obtaining  the ``experimental''  ANC values of astrophysical interest.   Besides, the proposed asymptotic  theory  can  also be applied to strong sub-barrier transfer reactions for which the main  contribution  to the reaction amplitude  comes to several  lowest partial waves $l_i$ ($l_i\sim\,k_iR_i^{{\rm{ch}}}$ =0, 1,..., where $k_i\,\to$  0 and $R_i^{{\rm{ch}}}\,\sim\, R_N$) and the contribution of   peripheral partial waves  $l_i$ ($l_i\,>>$ 1) is strongly suppressed.  

 It is worth  noting that the similar  theory   was   proposed earlier in  \cite{KMYa1988} for  the  peripheral    neutron transfer reaction induced by the   heavy ions at above-barrier energies,  which was also  implemented successfully for the specific  reactions.  However, for   peripheral charged particle transfer reactions   this task  requires  a special consideration. This is connected with  the considerable  complication occurring in  the main mechanisms  of the reaction reaction (\ref{subeq1}) because of   correct   taking into account  of   the three-body  Coulomb dynamics of the transfer mechanism   \cite{Av86,Mukh10}. 
   
   Below, we use the system of units  $c$= $\hbar$= 1 everywhere,   except where  they   are specially pointed out.
 
 \vspace{1cm}
 \begin{center}  
  {\bf II.  THREE-BODY COULOMB DYNAMICS OF THE TRANSFER MECHANISM    AND THE GENERALIZED    DWBA}
\end{center}
\bigskip

\hspace{0.6cm} We  consider the  reaction (\ref{subeq1}) within the framework of the     three ($A$, $a$ and $y$)   structureless  charged  particles. 
 In strict three-body  Schr\"{o}dinger approach, the amplitude for the reaction  (\ref{subeq1}) is given by  \cite{Gre1966,Austern1964}
\begin{equation} 
\label{subeq2}
 M^{{\rm{TB}}}{\rm{(}}E_i,\,cos\theta{\rm{)}}\,=\,\sum_{M_a}\langle\chi^{(-)}_{{\mathbf k}_f}I_{Aa}|V^{{\rm{TB}}}|I_{ay}\chi^{(+)}_{{\mathbf k}_i}\rangle
\end{equation}
and
\begin{equation}
\label{subeqTB}
V^{{\rm{TB}}}\,=\,\bigtriangleup V_f\,+\,\bigtriangleup V_fG\bigtriangleup V_i.
\end{equation}
 Here $\chi^{(+)}_{{\mathbf k}_i} $ and $ \chi^{(-)}_{{\mathbf k}_f}$ are the optical Coulomb--nuclear distorted wave functions in the entrance and exit channels   with the relative momentum  ${\mathbf  k}_i$ and ${\mathbf k}_f$, respectively ($E_i\,=\,k^2_i/2\mu_{Ax}$ and $E_f\,=\,k^2_f/2\mu_{By}$); $I_{Aa}$(${\mathbf r}_{Aa}$)($I_{ay}$(${\mathbf r}_{ay}$))  is  the overlap integral of the  bound-state      $\psi_A$, $\psi_a$ and $\psi_B$ ($\psi_y$, $\psi_a$ and $\psi_x$) wave functions \cite{Ber1965,BSat1965};     $\bigtriangleup V_f\,=\,V_{ay}\,+\,V_{yA}\,-\,V_f$;  $\bigtriangleup V_i\,=\,V_{Aa}\,+\,V_{yA}\,-\,V_i$;
 $G$ = (${\cal E}\, -\, H\,+\,i\cdot{\rm{o}}$)$^{-1}$ is  the operator of the three-body ($A$, $a$ and $y$)  Green's function  and $M_a$ is the spin  projections of the transferred particle $a$, where  $V_{ij}=V^N_{ij}+V^C_{ij}$, $V^N_{ij}$($V^C_{ij}$)  is the   nuclear (Coulomb) interaction  potential  between the centers  of mass of the  particles $i$ and $j$, which does not depend on the coordinates of the constituent nucleus;
$V_i$ and $V_f$ are the optical Coulomb--nuclear potentials in the entrance and  exit states, respectively;
  ${\cal E}\, =\, E_i\, -\, \varepsilon_{ay}\,=\, E_f\,-\,\varepsilon _{Aa}$ in which  $\varepsilon_{ij}$ is the binding energy of the  bound ($ij$)   system  in respect to    the    ($i\,+\,j$)  channel; ${\mathbf r}_{ij}\,=\, {\mathbf r}_{i}\,-\,{\mathbf r}_{j}$, ${\mathbf r}_{i}$ is the radius-vector of the center of mass of  the particle $i$ and $\mu_{ij}$ = $m_im_j/(m_i\,+\,m_j$ is the reduced mass of  the  $i$ and $j$   particles in which $m_j$ is the mass of the $j$ particle. 

 The operator of the three-body Green's function $G$  can be presented as
\begin{equation} \label{subeq3TB}
G\,=\,G_C\,+\,G_CV^NG,
\end{equation}
where $G_C$\,=\,(${\cal E}\, - \,T\, -\, V^C\,+\,i\cdot{\rm{0}}$)$^{-1}$   is  the operator of the three-body
($A$, $a$  and  $y$)  Coulomb   Green's functions;  $T$ is  the kinetic energy  operator for  the three-body
 ($A$, $a$ and $y$) system;  $V^N\,=\,V^N_{ay}\,+\,V^N_{Aa}\,+\,V^N_{yA}$ and  $V^C\,=\,V^C_{ay}\,+\,V^C_{Aa}\,+\,V^C_{yA}$.

The  overlap function $I_{Aa}{\rm{(}}{\mathbf r}_{Aa}{\rm{)}}$  is
given by \cite{Blok77} 
$$ I_{Aa}{\rm{(}}{\mathbf
r}_{Aa}{\rm{)}}\,=\,N_{Aa}^{1/2}\langle\psi_A{\rm{(}}\boldsymbol{\zeta}_A{\rm{)}}\psi_a{\rm{(}}\boldsymbol{\zeta}_a{\rm{)}}|\psi_B{\rm{(}}\boldsymbol{\zeta}_A,\boldsymbol{\zeta}_a;{\mathbf r}_{Aa}{\rm{)}}\rangle 
$$ 
\begin{equation} \label{subeq3TBOI}
=\,\sum_{l_B\mu_Bj_B\nu_B}C_{j_B\nu_BJ_AM_A}^{J_BM_B}C_{l_B\mu_B J_aM_a}^{j_B
 \nu_B}i^{l_B}Y_{l_B\mu_B}{\rm{(}}\hat{{\mathbf r}}_{Aa}{\rm{)}}I_{Aa;\,l_Bj_B}{\rm{(}}r_{Aa}{\rm{)}}.
  \end{equation}
Here $J_j(M_j)$ is the spin
(its projection) of the  particle $j$; $\hat{{\mathbf r}}_{Aa}={\mathbf r}_{Aa}/r_{Aa}$, $j_B$ and $\nu_B$
($l_B$ and $\mu_B$) are the total (orbital) angular momentum and its projection   of the particle $a$ in the nucleus $B$[=\,($A+a$)], respectively;
$C_{a\alpha\, b\beta }^{c\gamma}$ is the Clebsch-Gordan coefficient, and $N_{Aa}$
 is the factor taking into account the nucleons' identity \cite{Blok77}, which
is absorbed in the radial overlap function $I_{Aa;l_Bj_B}{\rm{(}}r_{Aa}{\rm{)}}$ being  not normalized to unity \cite{Ber1965}.   In the matrix element (\ref{subeq3TBOI}), the integration is taken over all the internal relative coordinates $\zeta_A$ and $\zeta_a$ for the $A$ and $a$ nuclei.

 The asymptotic
behavior   of $I_{Aa;l_Bj_B}{\rm{(}}r_{Aa}{\rm{)}}$  at  $r_{Aa}>r^{{\rm{(}}N{\rm{)}}}_{Aa}$  is given by the relation 
\begin{equation}
 I_{Aa;l_Bj_B}{\rm{(}}r_{Aa}{\rm{)}}\,\simeq\, C_{Aa;l_B\,j_B}\frac{W_{-\eta_B;\,l_B+1/2}{\rm{(}}2\kappa_{Aa}
r_{Aa}{\rm{)}}}{r_{Aa}}, \label{subeq4SFOI} \end{equation}
 where 
$W_{\alpha;\beta}{\rm{(}}r_{Aa}{\rm{)}}$ is the
Whittaker function, $\eta_B\,=\,z_Az_ae^2\mu_{Aa}/\kappa _{Aa}$ is
the Coulomb parameter for the
$B\,=\,(A\,+\,a)$ bound state, $\kappa _{Aa}\,=\,\sqrt{2\mu_{Aa}\varepsilon
_{Aa}}$,  
$r_{ij}^{{\rm{(}}N{\rm{)}}}$ is the nuclear interaction
radius between $i$ and $j$ particles in the  bound  ($i\,+\,j$) state
and $C_{Aa;\,l_Bj_B}$ is the ANC  for  $A\, + \,a\,\to\, B$,  which is related to the nuclear vertex constant $G_{Aa;\,l_Bj_B}$ for the virtual decay $B\,\to\,A\, + \,a$  as \cite{Blok77}
\begin{equation}
G_{Aa;l_Bj_B}=-i^{l_B+\eta_{Aa}}\frac{\sqrt{\pi}}{\mu_{Aa}}C_{Aa;l_Bj_B}\cdot
\label{subeq4CG}
\end{equation} 
 Eqs. (\ref{subeq3TBOI})--(\ref{subeq4SFOI})  and the expression for the matrix element $M_{Aa}{\rm{(}}{\mathbf{q}}_{Aa}{\rm{)}}$ for  the virtual decay $B\,\to\,A\,+\,a$, which is  given by Eq. (A1) in Appendix and related to the  overlap function $I_{Aa}{\rm{(}}{\mathbf r}_{Aa}{\rm{)}}$, hold for the  matrix element   $M_{ay}{\rm{(}}{\mathbf{q}}_{ay}{\rm{)}}$ of the virtual decay $x\,\to\,y\,+\,a$ and the overlap function $I_{ay}{\rm{(}}{\mathbf r}_{ay}{\rm{)}}$.  

The first ($V_{ay}$) and second ($V_{yA}$) terms,  entering the first term of the right-hand side (r.h.s.) of (\ref{subeqTB}),   correspond to the mechanisms described by the pole and triangle diagrams in Figs. \ref{fig1}$a$ and  \ref{fig1}$b$, respectively, where  the Coulomb-nuclear core-core ($A\,+\,y\,\longrightarrow\,A\,+\,y$) scattering in the  four-ray  vertex  in the  diagram in Fig. \ref{fig1}$b$ is taken  in the Born approximation. The $\bigtriangleup V_fG\bigtriangleup V_i$ term in the r.h.s. of (\ref{subeqTB}) corresponds to more complex mechanisms than the pole and triangle ones. This term  is described by a sum of nine diagrams obtained from the basic diagrams presented in   Figs. \ref{fig1}$a$ and  \ref{fig1}$b$,  which take into account  all possible subsequent mutual Coulomb-nuclear rescattering of the particles $A$, $a$ and $y$ in the intermediate state. One of the nine diagrams corresponding to the term $ V_{yA}GV_{Aa}$ is plotted in  Fig. \ref{fig1}$c$, where  the Coulomb-nuclear ($y\,+\,A\,\longrightarrow\,y\,+\,a$ and $A\,+\,a\,\longrightarrow\,A\,+\,a$) scatterings in the four-ray vertices, including in all four-ray vertices for the  others of   eight diagrams,   are taken in  the  Born approximation. This term corresponds to  the mechanism of subsequent Coulomb-nuclear rescattering of  the $y$ and $a$ particles, virtually emitted by the projectile $x$,  on the target $A$  in the intermediate state. In particular, it corresponds     to   the  mechanism    of      the  subsequent  rescatterings of  the proton ($p$) and neutron ($n$),   virtually emitted  by  the deuteron in the field of the $A$ target    in the nucleon transfer $A$($d$, $N$)$B$ reaction,  where  $N$ is a nucleon,   the transferred particle  is either $p$ or $n$  and  $B$\,=\,$A$\,+\,$N$. 

If the reaction (\ref{subeq1}) is peripheral, then its dominant mechanism, at least in the main peak of the angular distribution, corresponds to the pole diagram  in Fig. \ref{fig1}$a$ \cite{DDM1973,KMYa1988}. The amplitude of this diagram has the  singularity at $\cos\theta\,=\,\xi$, which is the nearest one to the physical (-1 $\le\,\cos\theta\,\le$ 1) region \cite{Shapiro,DDM1973} and is given by the expression
\begin{equation}
\label{subeqSB19}
 \xi\,=\,\frac{k_{i{\rm{1}}}^2\,+\,k_f^2\,+\,\kappa_{ay}^2}{2k_{i{\rm{1}}}k_f}\,=\,\frac{k_i^2\,+\,k_{f{\rm{1}}}^2\,+\,\kappa_{Aa}^2}{2k_{i }k_{f{\rm{1}}}},
\end{equation}
 where  $k_{i{\rm{1}}}\,=\,{\rm{(}}m_{y}/m_x{\rm{)}}k_i$ and $k_{f{\rm{1}}}\,=\,{\rm{(}}m_A/m_B{\rm{)}}k_f$.
However,   if    we   ignore   nuclear interactions   in   the second ($V_{yA}$)  and the third  ($V_f$) terms of the first $\bigtriangleup V_f$ term of   the  r.h.s. of (\ref{subeqTB})    as well as in the  $\bigtriangleup V_fG\bigtriangleup V_i$  one    with the help of the corresponding  replacement
$$ 
 V_{yA}\,\longrightarrow\,V_{yA}^C,\,\, V_f\,\longrightarrow\,V_f^C,\,\, 
 \bigtriangleup V_fG\bigtriangleup V_i\, \longrightarrow\,\bigtriangleup V_f^CG_C\bigtriangleup V_i^C,
$$  where $\bigtriangleup V_f^C\,=\,V_{ay}^C\,+\,V_{yA}^C\,-\,V_f^C$ and  $\bigtriangleup V_i^C\,=\,V_{Aa}^C\,+\,V_{yA}^C\,-\,V_i^C$,  
  then the amplitude $ M^{{\rm{TB}}}{\rm{(}}E_i,\,cos\theta{\rm{)}}$ can be presented in the form
\begin{equation}
 M^{{\rm{TB}}}{\rm{(}}E_i,\,cos\theta{\rm{)}}\,\approx\, M^{{\rm{TBDWBA}}}{\rm{(}}E_i,\,cos\theta{\rm{)}}\,=\,M^{{\rm{DWBA}}}_{{\rm{post}}}{\rm{(}}E_i,\,cos\theta{\rm{)}}\,+\, \bigtriangleup M^{{\rm{TBDWBA}}}{\rm{(}}E_i,\,cos\theta{\rm{)}}.
\label{subeqSB5}
\end{equation}  
Here 
 \begin{equation} 
M^{{\rm{DWBA}}}_{{\rm{post}}}{\rm{(}}E_i,\,cos\theta{\rm{)}}\,=\,\sum_{M_a}\langle\chi^{(-)}_{{\mathbf k}_f}I_{Aa}|V_{ay}\,+\,V_{yA}^C\,-\,V_f^C |I_{ay}\chi^{(+)}_{{\mathbf k}_i}\rangle
 \label{subeqSB6}
\end{equation}  
and
 \begin{equation} 
 \bigtriangleup M^{{\rm{TBDWBA}}}{\rm{(}}E_i,\,cos\theta{\rm{)}} \,=
  \,\sum_{M_a}\langle\chi^{(-)}_{{\mathbf k}_f}I_{Aa}|\bigtriangleup V_f^CG_C\bigtriangleup V_i^C|I_{ay}\chi^{(+)}_{{\mathbf k}_i}\rangle\cdot
\label{subeqSB7}
\end{equation}
 In Eqs (\ref{subeqSB5})--(\ref{subeqSB7}), the contribution of   the three-body ($A$, $a$ and $y$) Coulomb dynamics of  the transfer mechanism  in the intermediate state  involves    all orders of the perturbation theory over the optical  Coulomb polarization potential $\bigtriangleup V^C_{f,i}$, whereas  the Coulomb-nuclear distortions ($V_i$ and $V_f$) in the entrance and exit channels are  taken into account within the framework of  the optical model.  The amplitude  $M^{{\rm{TBDWBA}}}{\rm{(}}E_i,\,cos\theta{\rm{)}}$ can be considered as generalization of the ``post'' form of  the DWBA amplitude ($M^{{\rm{DWBA}}}_{{\rm{post}}}{\rm{(}}E_i,\,cos\theta{\rm{)}}$) \cite{LOLA} in which the three-body Coulomb dynamics of the main transfer mechanism are taken into account  in a correct manner. One notes that  the amplitude $ M^{{\rm{TBDWBA}}}{\rm{(}}E_i,\,cos\theta{\rm{)}}\,$ passes to the amplitude of the     so-called    ``post''-approximation of the  DWBA   if all the terms of $\bigtriangleup V_{f,i}^C$ contained in the transition operators of Eqs. (\ref{subeqSB6})  and (\ref{subeqSB7})   are ignored.

\begin{center}
\vspace{1cm}  
  {\bf III.   DISPERSION APPROACH AND   DWBA}
  \end{center}
  \bigskip

  The amplitudes given by  Eqs. (\ref{subeqSB6}) and (\ref{subeqSB7})  have   the  nearest singularity $\xi$ (the type of  branch point), which defines   the behavior  both of the  amplitude $ M^{{\rm{TB}}}{\rm{(}}E_i,\,cos\theta{\rm{)}}$ at $\cos\theta$ = $\xi$ \cite{Mukh10}  and   of    the   true  peripheral partial amplitudes at $l_i\,>>$ 1 \cite{Popov1964}.  Besides,    owing to   the presence of nuclear distortions in the entrance and exit states,  these amplitudes have also the singularities  located farther from the physical region than  $\xi$.  Nevertheless,    the behavior of the $M^{{\rm{DWBA}}}_{\rm{post}}{\rm{(}}E_i,\,cos\theta{\rm{)}}$  near  $\cos\theta\,=\,\xi$, denoted by  $M^{{\rm(}s{\rm{)}}\,DWBA}_{{\rm{post}}}{\rm{(}}E_i,\,cos\theta{\rm{)}}$ below, can be presented   as: 
\begin{equation} 
M^{{\rm(}s{\rm{)}}\,{\rm{DWBA}}}_{{\rm{post}}}{\rm{(}}E_i,\,cos\theta{\rm{)}}\,=\,{\cal{R}}^{{\rm{DWBA}}}_{{\rm{post}}}M^{{\rm(}s{\rm{)}}\,{\rm{DWBA}}}_{{\rm{pole}}}{\rm{(}}E_i,\,cos\theta{\rm{)}}, 
 \label{subeqSB8}
\end{equation} 
\begin{equation}{\cal{R}}^{{\rm{DWBA}}}_{{\rm{post}}}\,=\,\frac{N^{{\rm{DWBA}}}_{{\rm{post}}}}{N^{{\rm{DWBA}}}_{{\rm{pole }}}}\cdot 
\label{subeqSB8a}
\end{equation} 
Here $M^{{\rm(}s{\rm{)}\,DWBA}}_{{\rm{pole}}}{\rm{(}}E_i,\,cos\theta{\rm{)}}$ is the behavior of the  $ M^{{\rm{DWBA}}}_{{\rm{pole}}}{\rm{(}}E_i,\,cos\theta{\rm{)}} $ amplitude near  $\cos\theta\,=\,\xi$ \cite{Mukh10}, which corresponds to  the mechanism described by the diagram  in Fig. \ref{fig1}$a$ and is determined from  Eq. (\ref{subeqSB6})  if the   $V_{yA}^C\,-\,V_f^C$ term in the transition operator is ignored. In (\ref{subeqSB8a}),    $N^{{\rm{DWBA}}}_{{\rm{pole}}}$ is the Coulomb  renormalized  factor (CRF) for  the pole-approximation of the DWBA   amplitude    and  $N^{{\rm{DWBA}}}_{{\rm{post}}}$ is the CRF for the $M^{{\rm(}s{\rm{)}}\,{\rm{DWBA}}}_{{\rm{post}}}{\rm{(}}E_i,\,cos\theta{\rm{)}}$ amplitude.  The explicit forms of the CRFs  $N^{{\rm{DWBA}}}_{{\rm{pole}}} $ and $N^{{\rm{DWBA}}}_{{\rm{post}}}$  are given in Eqs. (14) and (26) of \cite{Mukh10}. As for,  the behavior  of the singular part of  the  $\bigtriangleup M^{{\rm{TBDWBA}}}{\rm{(}}E_i,\,cos\theta{\rm{)}}$ amplitude at $\cos\theta$ = $\xi$,  as is pointed out in   \cite{Mukh10}, it has the identical behaviour as that for 
 the $M_{{\rm{post}}}^{{\rm(}s{\rm{)}}\,{\rm{DWBA}}}{\rm{(}}E_i,\,cos\theta{\rm{)}}$ 
amplitude.  But,     the task of  directly  finding   the CRF explicit form  of   the $\bigtriangleup M^{{\rm{TBDWBA}}}{\rm{(}}E_i,\,cos\theta{\rm{)}}$  amplitude    is fairly difficult    because of the presence of the three-body Coulomb operator $G_C$ in the transition operator    and, so,   it  requires a special consideration, especially, in  the so-called ``dramatic'' case \cite{Mukh10}. In this case,   the partial wave amplitudes with $l_i>>$ 1 generating  the behavior of the $\bigtriangleup M^{{\rm{TBDWBA}}}{\rm{(}}E_i,\,cos\theta{\rm{)}}$ amplitude at $\cos\theta$ = $\xi$ provide the essential contribution  to the amplitude $ M^{{\rm{TB}}}{\rm{(}}E_i,\,cos\theta{\rm{)}}$ \cite{Mukh10}. For example, as it is noted  in \cite{Igam072}, the peripheral  ${\rm{^{10}B(^7Be,\,^8B)^9Be}}$ and ${\rm{^{14}N(^7Be,\,^8B)^{13}C}}$ reactions considered in  \cite{Azh1999,Azh1999-1,Tab06} within the ``post'' form of the MDWBA are related in the ``dramatic'' case. Perhaps, that is one of the possible reasons why the ANC value for ${\rm{^7Be}}\,+\,p\to{\rm{^8B}}$ recommended in  \cite{Tab06} is underestimated, which lead in turn to the underestimated astrophysical $S$ factor for the direct radiative capture ${\rm{^7Be(}}p,\,\gamma {\rm{)^8B}}$ reaction at solar energies \cite{YaBl2018}. The analogous case occurs for the peripheral ${\rm{^{14}N(^{13}N,\,^{14}O)^{13}C}}$ reaction, which was  analyzied in \cite{Tang2004} within the  same MDWBA. This fact dictates the further updating of the asymptotic theory proposed in the present work, where    the ``dramatic'' case  will also be included. At present such work  is in progress.

 Nevertheless, in the  ``non-dramatic'' case,  the accuracy of the  $M^{{\rm(}s{\rm{)}}\,DWBA}_{{\rm{post}}}{\rm{(}}E_i,\,cos\theta{\rm{)}}$ amplitude can be defined  by the extent of proximity of the CRF $N^{DWBA}_{{\rm{post}}}$ and  the true CRF $N^{TB}$ corresponding to  the $M^{{\rm{TB}}}{\rm{(}}E_i,\,cos\theta{\rm{)}}$ amplitude \cite{Mukh10} (see Table 2 in  \cite{Mukh10}).  The explicit form of $N^{TB}$ has been obtained in \cite{Av86} from the exact (in the framework of the three-body ($A$, $a$ and $y$) charged particle model)  amplitude of the sub-barrier reaction (\ref{subeq1}) and  is  given in  Refs. \cite{Av86,Mukh10}. Then the behavior of the exact   $ M^{{\rm{TB}}}{\rm{(}}E_i,\,cos\theta{\rm{)}}$ amplitude,  denoted by $ M^{{\rm{(}}s{\rm{)}}\,{\rm{TB}}}{\rm{(}}E_i,\,cos\theta{\rm{)}}$ below, near  the  singularity at $\cos\theta$ = $\xi$ takes the form as
\begin{equation} 
M^{ {\rm{TB}}}{\rm{(}}E_i,\,cos\theta{\rm{)}}\approx M^{{\rm(}s{\rm{)}}\,{\rm{TB}}}{\rm{(}}E_i,\,cos\theta{\rm{)}}\,=\,{\cal{R}}^{TB}M^{{\rm(}s{\rm{)}}\,{\rm{DWBA}}}_{{\rm{pole}}}{\rm{(}}E_i,\,cos\theta{\rm{)}}, 
 \label{subeqSB9}
\end{equation} where 
\begin{equation}
{\cal{R}}^{{\rm{TB}}}\,=\,\frac{N^{{\rm{TB}}}}{N^{{\rm{DWBA}}}_{{\rm{pole}}}}\cdot
\label{subeqSB10}
\end{equation}
One can see that  the  $M^{{\rm(}s{\rm{)}}\,{\rm{DWBA}}}_{{\rm{pole}}}{\rm{(}}E_i,\,cos\theta{\rm{)}}$,  $M^{{\rm(}s{\rm{)}}\,{\rm{DWBA}}}_{{\rm{post}}}{\rm{(}}E_i,\,cos\theta{\rm{)}}$ and   $M^{{\rm(}s{\rm{)}}\,{\rm{TB}}}{\rm{(}}E_i,\,cos\theta{\rm{)}}$ amplitudes 
 have the same behaviour near the singular point at  $\cos\theta$ = $\xi$. But,  they differ from each other only  by the power.  These amplitudes define   the corresponding  peripheral partial amplitudes for $l_i\,>>$ 1, which differ also from each other by their power \cite{Popov1964}. 
 
  Therefore, below we will first show how to obtain the singular part of the $M^{{\rm{DWBA}}}_{{\rm{pole}}}{\rm{(}}E_i,\,cos\theta{\rm{)}}$  by separating  the contribution  from the nearest singularity $\xi$ to it. Then, from the expression derived for this amplitude  we obtain the generalized DWBA amplitude in which  the contribution of  the three-body ($A$, $a$ and $y$) Coulomb dynamics of the main transfer mechanism to    the   peripheral partial amplitudes for  $l_i\,>>$ 1    are taken into account  in a correct manner.

\vspace{1cm} 
\begin{center} 
  {\bf IV.   DISTORTED-WAVE POLE APPROXIMATION}
  \end{center}
  \bigskip

 The pole-approximation of the  DWBA  amplitude  can  be obtained  from  Eq.
   (\ref{subeqSB6}). It has  the form as
 \begin{equation}
  M^{{\rm{DWBA}}}_{{\rm{pole}}}{\rm{(}}E_i,\,cos\theta{\rm{)}}\,=\,\int d{\mathbf{r}}_id{\mathbf{r}}_f \chi^{(-)^*}_{{\mathbf k}_f}({\mathbf{r}}_f)
I^*_{Aa}({\mathbf{r}}_{Aa}) V_{ay}({\mathbf{r}}_{ay}) I_{ay}({\mathbf{r}}_{ay}) \chi^{(+)}_{{\mathbf{ k}}_i}({\mathbf{r}}_i).
 \label{subeqSB13}
\end{equation}Here ${\mathbf{r}}_i\,\equiv\,{\mathbf{r}}_{xA} $,  ${\mathbf{r}}_f\,\equiv\,{\mathbf{r}}_{yB} $  and  
 $$
{\mathbf{r}}_{ay}\,=\,\bar{a}{\mathbf{r}_i}\,-\,\bar{b} {\mathbf{r}_f},
$$ 
 \begin{equation}
\label{subeqSB14}
 {\mathbf{r}}_{Aa}\,=\,-\,\bar{c}{\mathbf{r}_i}\,+\,\bar{d}{\mathbf{r}_f},
  \end{equation}where $\bar {a}$= $\mu_{Ax}/m_a$, $\bar{b}$= $\mu_{Ax}/\mu_{Aa}$, $\bar{c}$= $\mu_{By}/\mu_{ay}$   and $\bar{d}$= $\mu_{By}/m_a$. 
  To obtain the explicit  singular behavior of  $M^{{\rm{DWBA}}}_{{\rm{pole}}}{\rm{(}}E_i,\,cos\theta{\rm{)}}$ at $\cos\theta=\,\xi$, the integral (\ref{subeqSB13}) should be rewritten  in the momentum representation making use  of Eq. (A1) from  Appendix. It takes  the form 
\begin{equation}
  M^{{\rm{DWBA}}}_{{\rm{pole}}}{\rm{(}}E_i,\,cos\theta{\rm{)}}\,=\,\int\frac{ d{\mathbf{k}}}{(2\pi)^3}\frac{d{\mathbf{k}}^{\,\prime}}{(2\pi)^3} \chi^{(+)}_{{\mathbf k}_f}({\mathbf{k}}^{\,\prime})
{\cal{ M}}^{{\rm{DWBA}}}_{{\rm{pole}}}{\rm{(}}{\mathbf{k}}^{\,\prime},\,{\mathbf{k}}{\rm{)}} \chi^{(+)}_{{\mathbf{ k}}_i}({\mathbf{k}}),
 \label{subeqSB15}
\end{equation} 
$$
 {\cal{ M}}^{{\rm{DWBA}}}_{{\rm{pole}}}{\rm{(}}{\mathbf{k}}^{\,\prime},\,{\mathbf{k}}{\rm{)}}
 \,= \,\sum_{M_a}\langle {\mathbf{k}}^{\,\prime},\,I_{Aa}({\mathbf{q}}_{Aa})|V_{ay}({\mathbf{q}}_{ay})|
I_{ay}  ({\mathbf{q}}_{ay}),\,{\mathbf{k}}\rangle
$$
\begin{equation}
=\,-\,\sum_{M_a}\frac{M_{ay}{\rm{(}}{\mathbf{q}}_{ay}{\rm{)}} M_{Aa}^*{\rm{(}}{\mathbf{q}}_{Aa}{\rm{)}}}{\frac{q_{Aa}^{{\rm{2}}}}{{\rm{2}}\mu_{Aa}}\,+\,\varepsilon_{Aa}}
 \label{subeqSB16}
\end{equation} 
 Here ${\cal{ M}}^{{\rm{DWBA}}}_{{\rm{pole}}}{\rm{(}}{\mathbf{k}}^{\,\prime},\,{\mathbf{k}}{\rm{)}}$ is the off-shell   of the Born (pole) amplitude;     $\chi^{(+)}_{{\mathbf{ k}}_i}({\mathbf{k}})$  ($ \chi^{(+)}_{{\mathbf{ k}}_i}({\mathbf{k}})$), $I_{ay}({\mathbf{q}}_{ay})$ ($I_{Aa}({\mathbf{q}}_{Aa})$) and $V_{ay}({\mathbf{q}}_{ay})$    are   the Fourier components of the  distorted wave function  in the  entrance (exit) channel,    the overlap  function for the bound ($y$ + $a$) (($A$ + $a$)) state and  the Coulomb-nuclear $V_{ay}({\mathbf{r}}_{ay}) $ potential,  respectively;       ${\mathbf{q}}_{ay}\,=\,{\mathbf{k}}_1\,-\,{\mathbf{k}}^{\,\prime}$ and   ${\mathbf{q}}_{Aa}\,=\,-\,{\mathbf{k}}\,+\,{\mathbf{k}}^{\,\prime}_{{\rm{1}}}$, where  ${\mathbf{k}}_1\,=\,{\rm{(}}m_y/m_x{\rm{)}}{\mathbf{k}}$ and ${\mathbf{k}}^{\,\prime}_{{\rm{1}}}\,=\,{\rm{(}}m_A/m_B{\rm{)}}{\mathbf{k}}^{\,\prime}$. The explicit form of $M_{ay}{\rm{(}}{\mathbf{q}}_{ay}{\rm{)}}$   is  similar to that for the virtual decay $B\to\,A\,+\,a$ given by  Eq. (A1)   in Appendix.
 
  Using Eq. (A1) from Appendix and the  corresponding expression  for  $M_{ay}{\rm{(}}{\mathbf{q}}_{ay}{\rm{)}}$,   the  ${\cal{ M}}^{{\rm{DWBA}}}_{{\rm{pole}}}{\rm{(}}{\mathbf{k}}^{\,\prime},\,{\mathbf{k}}{\rm{)}} $ amplitude can be presented in  the form
$$
 {\cal{ M}}^{{\rm{DWBA}}}_{{\rm{pole}}}{\rm{(}}{\mathbf{k}}^{\,\prime},\,{\mathbf{k}}{\rm{)}}
 \,=\, \sum_{\alpha_B\alpha_xM_a}C{\rm{(}}\alpha_B\alpha_x;\,{\rm{(}}J,\,M{\rm{)}}_{x,\,A,\,y,\,B};\,J_aM_a{\rm{)}}
 {\tilde{{\cal{ M}}}}^{{\rm{DWBA}}}_{{\rm{pole}};\,\alpha_B\alpha_x}{\rm{(}}{\mathbf{k}}^{\,\prime},\,{\mathbf{k}}{\rm{)}},
$$
\begin{equation}
{\tilde{{\cal{ M}}}}^{{\rm{DWBA}}}_{{\rm{pole}};\,\alpha_B\alpha_x}{\rm{(}}{\mathbf{k}}^{\,\prime},\,{\mathbf{k}}{\rm{)}}\,=\,I^{^*}_{Aa;\,\alpha_B}{\rm{(}}{\mathbf{q}}_{Aa}{\rm{)}}W_{ay;\,\alpha_x}{\rm{(}}{\mathbf{q}}_{ay}{\rm{)}}.  
 \label{subeqSB17}
\end{equation} Here 
$$
C{\rm{(}}\alpha_B\alpha_x;\,{\rm{(}}J,\,M{\rm{)}}_{x,\,A,\,y,\,B};\,J_aM_a{\rm{)}}\,=\,C_{j_x\nu_xJ_yM_y}^{J_xM_x}C_{l_x\mu_xJ_aM_a}^{j_x\nu_x}C_{j_B\nu_BJ_AM_A}^{J_BM_B}
C_{l_B\nu_BJ_aM_a}^{j_B\nu_B} 
$$  and
\begin{equation}
I_{Aa;\,\alpha_B}{\rm{(}}{\mathbf{q}}_{Aa}{\rm{)}}=\,-\,{\rm{2}}\mu_{Aa}\frac{W_{Aa;\,\alpha_B}{\rm{(}}{\bf q}_{ya}{\rm{)}}}{q_{Aa}^{{\rm{2}}}+\kappa_{Aa}^{{\rm{2}}}},
\label{subeqSB17a}
\end{equation}  where $\alpha_{\lambda}\,=\,{\rm{(}}l_{\lambda},\,\mu_{\lambda},\,j_{\lambda},\,\nu_{\lambda}{\rm{)}}$;  $\lambda\,=\,x,\,B$; ($J,\,M$) is the set of $J_{\lambda}$ and $M_{\lambda}$ ($\lambda\,=\,x,\,A,\,y,\,B$)   and
$$
W_{Aa;\,\alpha_B}{\rm{(}}{\bf q}_{Aa}{\rm{)}}=\,\sqrt{{\rm{4}}\pi}G_{Aa;\,l_Bj_B}{\rm{(}}q_{Aa}{\rm{)}}Y_{l_B\mu_B}(\hat{{\mathbf q}}_{Aa}{\rm{)}},
$$  
\begin{equation}
W_{ay;\,\alpha_x}{\rm{(}}{\mathbf{q}}_{ay}{\rm{)}}=\,\sqrt{{\rm{4}}\pi}G_{ay;\,l_xj_x}{\rm{(}}q_{ay}{\rm{)}}Y_{l_x\mu_x}(\hat{{\mathbf q}}_{ay}{\rm{)}}
\label{subeqSB17b}
\end{equation} are     the reduced  vertex functions for the virtual decays  $B\,\to\,A\,+\,a$  and $x\,\to\,\,y\,+\,a$, respectively.
  
\begin{figure}
\begin{center}
\includegraphics[height=4cm, width=14cm]{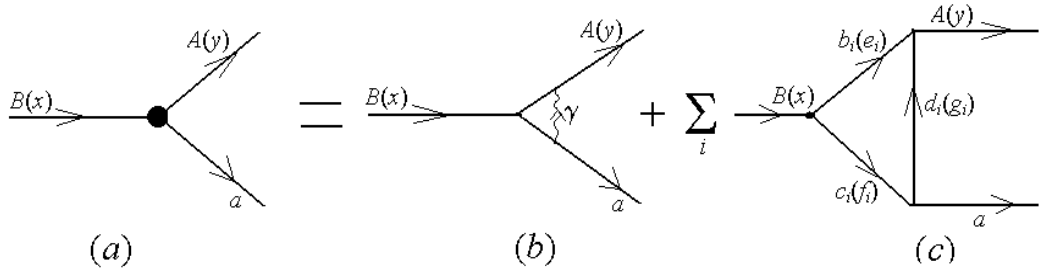}
 \end{center}
\caption{\label{fig2} Diagrams describing   the matrix element for  the virtual decay $B\,\to\,A\,+\,a$ ($x\,\to\,y\,+\,a$).}
\end{figure}

  In the presence of the long-range Coulomb interactions between particles $A$ and $a$ ($y$ and $a$), the reduced  vertex function     for the virtual decay $B\,\to\,A\,+\,a$ ($x\,\to\,\,y\,+\,a$)  can be described by the sum of the nonrelativistic diagrams plotted in Fig. \ref{fig2}. The diagram  in Fig. \ref{fig2}$b$ corresponds  to the Coulomb part of the corresponding  vertex function, which  has a branch point singularity at the point $q^{{\rm{2}}}_{Aa}+\kappa^{{\rm{2}}}_{Aa}$ =0 ($q^{{\rm{2}}}_{ay}\,+\,\kappa^{{\rm{2}}}_{ay}$ = 0) and   generates    the singularity of the $  M^{{\rm{DWBA}}}_{{\rm{pole}}}{\rm{(}}E_i,\,cos\theta{\rm{)}}$ amplitude  at  $\cos\theta$ = $\xi$. The sum in Fig. \ref{fig2}$c$ involves more complicated diagrams  and this part  of the vertex function  corresponds to the   Coulomb-nuclear vertex function, which   is regular at the point $q_{Aa}\,=\,i\kappa_{Aa}$  ($q_{ay}\,=\,i\kappa_{ay}$).  Then, the vertex functions  $W_{Aa;\,\alpha_B}{\rm{(}}{\bf q}_{Aa}{\rm{)}}$  and $W_{ay;\,\alpha_x}{\rm{(}}{\mathbf{q}}_{ay}{\rm{)}} $ can be presented in the forms \cite{DDh1971}
  $$
  W_{Aa;\,\alpha_B}{\rm{(}}{\mathbf q}_{Aa}{\rm{)}}=
 W^{{\rm{(C)}}}_{ Aa;\,\alpha_B}{\rm{(}}{\mathbf q}_{Aa}{\rm{)}}\,+\,
W^{{\rm{(CN)}}}_{Aa;\,\alpha_B}{\rm{(}}{\mathbf q}_{Aa}{\rm{)}},\,\,\,
  $$
\begin{equation}
W_{ay;\,\alpha_x}{\rm{(}}{\bf q}_{ay}{\rm{)}}=\,
W^{{\rm{(C)}}}_{ay;\,\alpha_x}{\rm{(}}{\mathbf q}_{ay}{\rm{)}}\,+\,
W^{{\rm{(CN)}}}_{ya;\,\alpha_x}{\rm{(}}{\mathbf q}_{ay}{\rm{)}}.
\label{subeqSB17WBx}
\end{equation} Here $W^{{\rm{(C)}}}_{Aa;\,\alpha_B}$ and $W^{{\rm{(CN)}}}_{Aa;\,\alpha_B}$
($W^{{\rm{(C)}}}_{ay;\,\alpha_x}$ and $W^{{\rm{(CN)}}}_{ay;\,\alpha_x}$) are the Coulomb part and the regular function at the point    $q_{Aa}\,=\,i\kappa_{Aa}$ ($q_{ay}\,=\,i\kappa_{ay}$), respectively.
   All terms of the sum in  Fig. \ref{fig2}$c$  have   dynamic singularities, which are  generated     by internuclear interactions being responsible for   taking into account of the so-called  dynamic recoil effects \cite{Austern1964,LOLA}.  These singularities are   located   at the points  $q_{Aa}\,=\,i\lambda_i\kappa_i$ and $q_{ay}\,=\,i\bar{\lambda}_i\bar{\kappa}_i$  \cite{BDP1963,Blokh2013}, where $\lambda_i\,=\,m_A/m_{b_i}$,  $\kappa_i=\,  \kappa_{b_ic_i}\,+\,\kappa_{b_id_i}$,  $\bar{\lambda}_i\,=\,m_y/m_{e_i}$ and $\bar{\kappa}_i=\, \kappa_{e_if_i}\,+\,\kappa_{e_ig_i} $.  They   generate  the singularities $\xi_i$ and $\bar{\xi}_ i $ of the $  M^{{\rm{DWBA}}}_{{\rm{pole}}}{\rm{(}}E_i,\,cos\theta{\rm{)}}$ amplitude, which are determined by   
$$
\xi_i=\,\frac{{\rm{(}}k_im_{b_i}/m_A{\rm{)}}^{{\rm{2}}}\,+\,{\rm{(}}k_fm_{b_i}/m_B{\rm{)}}^{{\rm{2}}}\,+\,\kappa_i^{{\rm{2}}}}{{\rm{2}}k_ik_fb^{{\rm{2}}}_i/m_Am_B}
$$ and 
$$
\bar{\xi}_i=\,\frac{{\rm{(}}k_im_{e_i}/m_x{\rm{)}}^{{\rm{2}}}\,+\,{\rm{(}}k_fm_{e_i}/m_y{\rm{)}}^{{\rm{2}}}\,+\,\bar{\kappa}_i^{{\rm{2}}}}{{\rm{2}}k_ik_fe_i^{{\rm{2}}}/m_xm_y}.
$$  
As a rule, they   are   located  farther from the physical (-1 $\le\,\cos\theta\,\le$ 1) region than $\xi$ ($\xi_i\,>\,\xi$ and $\bar{\xi}_i\,>\,\xi$) \cite{DDh1971,BDP1963}. For illustration of the fact, the  position of these singularities ($\xi$, $\xi_i$ and $\bar{\xi}_i$), $\kappa$, $\kappa_i$ and $\bar{\kappa}_i$ calculated  for the  specific  peripheral reactions considered in the present work   are presented in Table {\ref{table1}}. As can be  seen from Table \ref{table1}, the singularities $\xi_i$ and $\bar{\xi}_i$    are located farther from the physical (-1$\le\,\cos\theta\,\le$ 1) region  than the singularity $\xi$.
   
     For the surface reaction (\ref{subeq1}), the contribution of the interior nuclear range to the $M^{{\rm{DWBA}}}_{{\rm{pole}}}{\rm{(}}E_i,\,cos\theta{\rm{)}} $ amplitude, which is generated  by the singularities of the $W^{{\rm{(CN)}}}_{Aa;\,\alpha_B}$ and  $W^{{\rm{(CN)}}}_{ay;\,\alpha_x}$ functions,  can be ignored at least in the main peak of the angular distribution \cite{DDM1973,KMYa1988}.   Therefore,      the vertex functions    for the virtual decays $B\,\to\,A\,+\,a$  and $x\,\to\,\,y\,+\,a$ given  by Eq. (\ref{subeqSB17b})  can be replaced by their  Coulomb parts  in the vicinity of    nearest singularities (the branch points) located  at  
   $q_{ay}\,=\,i\kappa_{ay}$  and $q_{Aa}\,=i\kappa_{Aa}$, respectively. They    behave as \cite{DDh1971}
$$
 W^{{\rm{(C)}}}_{\beta\gamma;\,\alpha_{\alpha}}{\rm{(}}{\mathbf q}_{\beta\gamma}{\rm{)}} 
  \,\simeq\, W^{{\rm{(C}};\,s{\rm{)}}}_{\beta\gamma;\,\alpha_{\alpha}}{\rm{(}}{\mathbf q}_{\beta\gamma}{\rm{)}}
    =\,\sqrt{{\rm{4}}\pi}\Gamma{\rm{(}}1\,-\,\eta_{\beta\gamma}{\rm{)}}
$$
 \begin{equation}
\label{subeqSB18}    
\times\Big(\frac{q_{\beta\gamma}}{i\kappa_{\beta\gamma}}\Big)^{l_{\alpha}}
\Big(\frac{{\mathbf{q}}^{\rm{2}}_{\beta\gamma}
\,+\,\kappa_{\beta\gamma}^{\rm{2}}}
{{\rm{4}}i\kappa_{\beta\gamma}^{\rm{2}}}\Big)^{\eta_{\beta\gamma}}
G_{\beta\gamma;\,\,l_{\alpha}j_{\alpha}}
{\rm{(}}i\kappa_{\beta\gamma}{\rm{)}}Y_{l_{\alpha}\nu_{\alpha}}{\rm{(}}\hat {\bf q}_{\beta\gamma}{\rm{)}} 
\end{equation} for  $q_{\beta\gamma}\,\to\,
i\kappa_{\beta\gamma}$,  where $G_{\beta\gamma;\,\,l_{\alpha}j_{\alpha}}{\rm{(}}i\kappa_{\beta\gamma}{\rm{) (}}\equiv G_{\beta\gamma;\,\,l_{\alpha}j_{\alpha}}$) is the NVC for the virtual decay $\alpha\,\to\,\beta +\gamma$; $\gamma$ = $a$;    $\alpha$ = $x$   and $\beta$ = $y$ for the virtual decay $x\,\to\,y\,+\,a$, whereas     $\alpha$ = $B$   and $\beta$ = $A$ for the virtual decay   $B\,\to\,A\,+\,a$.    

As is seen from  Eqs. (\ref{subeqSB17}), (\ref{subeqSB17a}) and  (\ref{subeqSB18}), the off-shell Born amplitude ${\cal{ M}}^{{\rm{DWBA}}}_{{\rm{pole}}}{\rm{(}}{\mathbf{k}}^{\,\prime},\,{\mathbf{k}}{\rm{)}}$ at ${\mathbf{k}} $ = ${\mathbf{k}}_i$ and  ${\mathbf{k}}^{\,\prime}$ = ${\mathbf{k}}_f$ has the  nearest dynamic singularity at $\cos\theta$ = $\xi$.
 Besides,  ${\cal{ M}}^{{\rm{DWBA}}}_{{\rm{post}}}{\rm{(}}{\mathbf{k}}^{\,\prime},\,{\mathbf{k}}{\rm{)}}$  has also kinematic singularities generated by   the factors 
${\rm{(}}q_{ay}/i\kappa_{ay}{\rm{)}}^{l_x}Y_{l_x\mu_x}{\rm{(}}{\hat{{\mathbf{q}}}}_{ay}{\rm{)}}$ and $
{\rm{(}}q_{Aa}/i\kappa_{Aa}{\rm{)}}^{l_B} Y_{l_B\mu_B}^*{\rm{(}}{\hat{{\mathbf{q}}}}_{Aa}{\rm{)}}$ in 
 (\ref{subeqSB17}) \cite{DDM1973}.  Nevertheless, we take into account the contribution generated by  the kinematic singularities to  the  ${\cal{ M}}^{{\rm{DWBA}}}_{{\rm{pole}}}{\rm{(}}{\mathbf{k}}^{\,\prime},\,{\mathbf{k}}{\rm{)}} $ amplitude. Then  ${\tilde{{\cal{ M}}}}^{{\rm{DWBA}}}_{{\rm{pole}};\,\alpha_B\alpha_x}{\rm{(}}{\mathbf{k}}^{\,\prime},\,{\mathbf{k}}{\rm{)}}$ given by (\ref{subeqSB17}) in the approximation (\ref{subeqSB18}) takes the form
 \begin{equation}
\label{subeqSB19}
{\tilde{{\cal{ M}}}}^{{\rm{DWBA}}}_{{\rm{pole}};\,\alpha_B\alpha_x}{\rm{(}}{\mathbf{k}}^{\,\prime},\,{\mathbf{k}}{\rm{)}}\,\approx\,{\tilde{{\cal{ M}}}}^{{\rm{(}}s{\rm{)}};\,{\rm{DWBA}}}_{{\rm{pole}};\,\alpha_B\alpha_x}{\rm{(}}{\mathbf{k}}^{\,\prime},\,{\mathbf{k}}{\rm{)}}\,=\,I^{^*(s)}_{Aa;\,\alpha_B}{\rm{(}}{\mathbf{q}}_{Aa}{\rm{)}}W_{ay;\,\alpha_x}^{(s)}{\rm{(}}{\mathbf{q}}_{ay}{\rm{)}},  
\end{equation} where 
 \begin{equation} \label{subeqSB20x}
  W_{ay;\,\alpha_x}^{{\rm{(}}s{\rm{)}}}{\rm{(}}{\bf q}_{ay}{\rm{)}}=\,\sqrt{{\rm{4}}\pi}G_{ay;\,l_xj_x} \Gamma{\rm{(}}1-\eta_{ay}{\rm{)}}\Big(\frac{q_{ay}}{i\kappa_{ay}}\Big)^{l_x}\Big(\frac{q_{ay}^{{\rm{2}}}+\kappa_{ay}^{{\rm{2}}}}{{\rm{4}}i\kappa^{{\rm{2}}}_{ay}}\Big)^{\eta_{ay}}Y_{l_x\nu_x}(\hat {\bf q}_{ay}), 
\end{equation}
 \begin{equation} \label{subeqSB20}
  I_{Aa;\,\alpha_B}^{*\,{\rm{(}}s{\rm{)}}}{\rm{(}}{\bf q}_{Aa}{\rm{)}}=\,-\sqrt{4\pi}G_{Aa;\,l_Bj_B}  \Gamma{\rm{(}}1\,-\,\eta_{Aa}{\rm{)}}\left(\frac{q_{Aa}}{i\kappa_{Aa}}\right)^{l_B}\left(\frac{q_{Aa}^{{\rm{2}}}\,+\,\kappa_{Aa}^{{\rm{2}}}}{{\rm{4}}i\kappa^{{\rm{2}}}_{Aa}}\right)^{\eta_{Aa}}\frac{{\rm{2}}\mu_{Aa}}{q_{Aa}^{{\rm{2}}}\,+\,\kappa_{Aa}^{{\rm{2}}}}Y^*_{l_B\nu_B}{\rm{(}}\hat {\mathbf q}_{Aa}{\rm{)}}. 
\end{equation}

We  now rewrite the  integral  (\ref{subeqSB15})  taking into account 
Eqs.  (\ref{subeqSB19}) -- (\ref{subeqSB20}) in the coordinate representation. First, we  consider this presentation for  the Fourier components of the $ W_{ay;\,\alpha_x}^{{\rm{(}}s{\rm{)}}}{\rm{(}}{\bf q}_{ay}{\rm{)}}$ and $I_{Aa;\,\alpha_B}^{*\,{\rm{(}}s{\rm{)}}}{\rm{(}}{\bf q}_{Aa}{\rm{)}}$ functions:
\begin{equation}
\label{subeqSB21}
W^{{\rm{(}}as{\rm{)}}}_{x;\,\alpha_x}{\rm{(}}{\mathbf{r}}_{ay}{\rm{)}}=\int\frac{d{\mathbf{q}}_{ay}}{{\rm{(}}{\rm 2}\pi)^{{\rm{3}}}}e^{i{\mathbf{r}}_{ay}{\mathbf{q}}_{ay}}W^{{\rm{(}}s{\rm{)}}}_{x;\,\alpha_x}{\rm{(}}{\mathbf{q}}_{ay}{\rm{)}}  
\end{equation}
and
\begin{equation}
\label{subeqSB22}
I^{{\rm{(}}as{\rm{)}}}_{B;\,\alpha_B}({\mathbf{r}}_{Aa}{\rm{)}}=\int\frac{d{\mathbf{q}}_{Aa}}{{\rm{(}}{\rm 2}\pi)^{{\rm{3}}}}e^{i{\mathbf{r}}_{Aa}{\mathbf{q}}_{Aa}}I^{(s)}_{Aa;\,\alpha_B}({\mathbf{q}}_{Aa})  
\end{equation} 
Substituting     Eq. (\ref{subeqSB20x})  in Eq. (\ref{subeqSB21}) and Eq.  (\ref{subeqSB20}) in Eq.   (\ref{subeqSB22}),    the integration over the angular variables   can immediately   be  performed  making use  of the    expansion
$$
e^{i{\mathbf{q}}{\mathbf{r}}}=
\,{\rm{4}}\pi\sum_{l\nu}i^lj_l{\rm{(}}qr{\rm{)}}
Y^*_{l\nu}{\rm{(}}\hat {\mathbf q}{\rm{)}}Y_{l\nu}{\rm{(}}\hat {\mathbf r}{\rm{)}},
$$ where  $j_l{\rm{(}}z{\rm{)}}$ is a  spherical Bessel function \cite{Abr1970}. The  remaining integrals in $q_{ay}$  and $q_{Aa}$ can be done   with the use of formula 6.565(4)   and  Eq. (91) from   Refs. \cite{GR1980} and  \cite{Blok77}, respectively.  As a result, one obtains 
\begin{equation}
\label{subeqSB23}
 W^{{\rm{(}}as{\rm{)}}}_{ay;\,\alpha_x}{\rm{(}}{\mathbf{r}}_{ay}{\rm{)}}= -\frac{\sqrt{{\rm {2}}}\eta_{ay}}{\pi}G_{ay;\,l_xj_x} \left(\frac{\kappa_{ay}}{r_{ay}}\right)^{3/2}\frac{K_{l_x+{\rm{3/2}}+\eta_{ay}}{\rm{(}}\kappa_{ay}r_{ay}{\rm{)}}}{{\rm{(}}2i\kappa_{ay}r_{ay}{\rm{)}}^{\eta_{ay}}}i^{-l_x}Y_{l_x\nu_x}{\rm{(}}\hat{{\mathbf{r}}}_{ay}{\rm{)}}
\end{equation}
 for $r_{ay}\gtrsim R_x $ and
\begin{equation}
\label{subeqSB24}
I^{*{\rm{(}}as{\rm{)}}}_{Aa;\,\alpha_B}{\rm{(}}{\mathbf{r}}_{Aa}{\rm{)}}\,=\,-\frac{\sqrt{{\rm {2}}}}{\pi}
G_{Aa;\,l_Bj_B}\left(\frac{\mu_{Aa}^{{\rm{2}}}\kappa_{Aa}}{r_{Aa}}\right)^{{\rm{1/2}}}\frac{K_{l_B+{\rm{1}}/{\rm{2}}+\eta_{Aa}}{\rm{(}}\kappa_{Aa}r_{Aa}{\rm{)}}}{{\rm{(2}}i\kappa_{Aa}r_{Aa}{\rm{)}}^{\eta_{Aa}}}i^{-l_B}Y^*_{l_B\nu_B}{\rm{(}}\hat{{\mathbf{r}}}_{Aa}{\rm{)}}
 \end{equation}
for $r_{Aa}\gtrsim R_B$. Here $K_{{\tilde{\nu}}}{\rm{(}}z{\rm{)}}$ is a modified Hankel function \cite{Abr1970} and  $R_C=r_0C^{\rm{1/3}}$ is the radius of $C$ nucleus, where    $C$  is a   mass number of the  $C$ nucleus.   Using formula 9.235 (2)  from \cite{GR1980} and the relation  (\ref{subeq4CG}), the   leading asymptotic  terms of Eqs. (\ref{subeqSB23}) and (\ref{subeqSB24}) can be reduced  to the forms 
\begin{equation}
\label{subeqSB23as} 
 W^{{\rm{(}}as{\rm{)}}}_{ay;\,\alpha_x}{\rm{(}}{\mathbf{r}}_{ay}{\rm{)}}\,\approx\, V_{ay}^C{\rm{(}}r_{ay}{\rm{)}}I^{{\rm{(}}as{\rm{)}}}_{ay;\,\alpha_x}{\rm{(}}r_{ay}{\rm{)}}Y_{l_x\nu_x}{\rm{(}}\hat{{\mathbf{r}}}_{ay}{\rm{)}},\,
 \end{equation}
  for $r_{ay}\gtrsim R_x $ and
 \begin{equation}
\label{subeqSB24Bas}
I^{^*{\rm{(}}as{\rm{)}}}_{Aa;\,\alpha_B}{\rm{(}}{\mathbf{r}}_{Aa}{\rm{)}}\,\approx\,C_{l_Bj_B}
\frac{\exp\{-\kappa_{Aa}r_{Aa}-\eta_{Aa}\ln{\rm{{(2}}}\kappa_{Aa}r_{Aa}{\rm{)}}\}}{r_{Aa}}
Y^*_{l_B\nu_B}{\rm{(}}\hat{{\mathbf{r}}}_{Aa}{\rm{)}},
 \end{equation} for $r_{Aa}\gtrsim R_B$. In  (\ref{subeqSB23as}),  $V_{ay}^C{\rm{(}}r_{ay}{\rm{)}}=\, Z_aZ_ye^2/r_{ay}$ is the Coulomb interaction potential  between the centers  of mass of particles $y$ and $a$, and 
\begin{equation}
\label{subeqSB25xas} 
 I^{{\rm{(}}as{\rm{)}}}_{ay;\,\alpha_x}{\rm{(}}r_{ay}{\rm{)}}\,=\,
C_{l_xj_x}
\frac{\exp\{-\kappa_{ay}r_{ay}-\eta_{ay}\ln{\rm{{(2}}}\kappa_{ay}r_{ay}{\rm{)}}\}}{r_{ay} },
 \end{equation} which coincides with the leading term of the asymptotic behavior of the radial component of the overlap function $I_{ay}{\rm{(}}{\mathbf{r}}_{ay}{\rm{)}} \,\approx\,I^{{\rm{(}}as{\rm{)}}}_{ay;\,\alpha_x}{\rm{(}}r_{ay}{\rm{)}} Y_{l_x\nu_x}{\rm{(}}\hat{{\mathbf{r}}}_{ay}{\rm{)}}$  for $r_{ay}> R_x $. 
 
 Following by the way of \cite{Blokh2013}, it can  be shown that the leading terms of   the asymptotic expressions for 
 the radial components of the Coulomb-nuclear parts of  $W_{ay}{\rm{(}}\mathbf{r}_{ay}{\rm{)}}$ and $I_{Aa}{\rm{(}}\mathbf{r}_{Aa}{\rm{)}}$ (Fig.\ref{fig2}$c$), which are   generated by the singularities of $\xi_i$ and $\bar{\xi}_i$, respectively,  behave as 
 \begin{equation}
\label{subeqSB26xBNCas} 
W^{{\rm{(}CN)}}_{l_xj_x}{\rm{(}}r_{ay}{\rm{)}}\approx\,\sum_iW^{{\rm{(CN}};\,as{\rm{)}}}_{l_xj_x;\,i}{\rm{(}}r_{ay}{\rm{)}}, \,\,\,I^{{\rm{(}CN)}}_{l_Bj_B}{\rm{(}}r_{Aa}{\rm{)}}\approx\,\sum_iI^{{\rm{(CN};\,as{\rm)}}}_{l_Bj_B;\,i}{\rm{(}}r_{Aa}{\rm{)}}.
 \end{equation} Here 
 \begin{equation}
\label{subeqSB27xCNas} 
 W^{{\rm{(CN}};\,as{\rm{)}}}_{l_xj_x;\,i}{\rm{(}}r_{ay}{\rm{)}}=\,\bar{C}^{{\rm{(\,}}i{\rm{\,)}}}_{l_xj_x}
 \frac{\exp\{-[\bar{\kappa}_ir_{ay}
 \,+\,\eta_{e_if_i}\ln{\rm{(2}}\bar{\lambda}_i\kappa_{e_if_i}r_{ay}{\rm{)}}\,+\,\eta_{e_ig_i}\ln{\rm{(2}}\bar{\lambda}_i\kappa_{e_ig_i}r_{ay}{\rm{)}}]\}}{r_{ay}^{{\rm{2}}}}, 
  \end{equation} 
  \begin{equation}
\label{subeqSB27BCNas} 
 I^{{\rm{(CN}};\,as{\rm{)}}}_{l_Bj_B;\,i}{\rm{(}}r_{Aa}{\rm{)}}=\,C^{{\rm{(\,}}i{\rm{\,)}}}_{l_Bj_B}
 \frac{\exp\{-[\kappa_ir_{Aa}
 \,+\,\eta_{b_ic_i}\ln{\rm{(2}}\lambda_i\kappa_{b_ic_i}r_{Aa}{\rm{)}}\,+\,\eta_{b_id_i}\ln{\rm{(2}}\lambda_i\kappa_{b_id_i}r_{Aa}{\rm{)}}]\}}{r_{Aa}^{{\rm{2}}}}. 
  \end{equation} 
   Explicit expressions for $\bar{C}^{{\rm{(\,}}i{\rm{\,)}}}_{l_xj_x}$ and $C^{{\rm{(\,}}i{\rm{\,)}}}_{l_Bj_B}$ can be  obtained from Eqs. (A.4) and (A.5) of \cite{Blokh2013}, which are expressed in the terms of the product of the ANCs for the tri-rays vertices of the diagrams in  Fig. \ref{fig2}$c$. As is seen from the expressions  (\ref{subeqSB26xBNCas}), (\ref{subeqSB27xCNas})  and (\ref{subeqSB27BCNas}), if $\kappa_i\,>\,\kappa_{Aa}$ and 
  $\bar{\kappa}_i\,>\,\kappa_{ya}$, which occur for the peripheral  reactions presented in Table \ref{table1},  then the  asymptotic  terms generated by the singularities $\bar{\xi}_i$  and $\xi_i$ of  the $  M^{{\rm{DWBA}}}_{{\rm{pole}}}{\rm{(}}E_i,\,cos\theta{\rm{)}}$  amplitude decrease more rapidly with increasing $r_{ay}$ and $r_{Aa}$, respectively, than those of (\ref{subeqSB23as}) and (\ref{subeqSB24Bas}) generated by the singularity $\xi$.
 Therefore,   the use of the pole approximation is reasonable in calculations of the leading terms of  the peripheral partial wave amplitudes at $l_i\,>>$ 1. They correctly  give the dominant contribution to the  $M^{{\rm{DWBA}}}_{{\rm{pole}}}{\rm{(}}E_i,\,cos\theta{\rm{)}}$ at least in the main peak of the angular distribution and are correctly  determined by only the nearest singularity $\xi$ \cite{Popov1964}.

  Thus,  usage of  the pole approximation in the amplitude (\ref{subeqSB13})   is equivalent to the replacements of  $V_{ay}({\mathbf{r}}_{ay}) I_{ay}({\mathbf{r}}_{ay})$ and  $ I^{^*}_{Aa;\,\alpha_B}{\rm{(}}{\mathbf{r}}_{Aa}{\rm{)}} 
$   by $ W^{{\rm{(}}as{\rm{)}}}_{ay;\,\alpha_x}{\rm{(}}{\mathbf{r}}_{ay}{\rm{)}}$ and $I^{^*{\rm{(}}as{\rm{)}}}_{Aa;\,\alpha_B}{\rm{(}}{\mathbf{r}}_{Aa}{\rm{)}}$ in the integrand of Eq. (\ref{subeqSB13}), respectively. In this case,   in the coordinate representation,  
    the  $M^{{\rm{DWBA}}}_{{\rm{pole}}}{\rm{(}}E_i,\,cos\theta{\rm{)}}$ amplitude can be reduced  to the form as
$$
M^{{\rm{DWBA}}}_{{\rm{pole}}}{\rm{(}}E_i,\,cos\theta{\rm{)}}\simeq\, M^{{\rm(}s{\rm{)}}\,{\rm{DWBA}}}_{{\rm{pole}}}{\rm{(}}E_i,\,cos\theta{\rm{)}}=\,\sum_{\alpha_B\alpha_xM_a}
C{\rm{(}}\alpha_B\,\alpha_x;\,{\rm{(}}J,\,M{\rm{)}}_{x,\,A,\,y,\,B};\,J_aM_a{\rm{)}}
$$
\begin{equation}
\label{subeqSB25}
\times{\tilde{M}}^{{\rm{DWBA}}}_{{\rm{pole};\,\alpha_B\alpha_x}}{\rm{(}}E_i,\,cos\theta{\rm{)}},
\end{equation}where
\begin{equation}
\label{subeqSB26}
{\tilde{M}}^{{\rm{DWBA}}}_{{\rm{pole};\,\alpha_B\alpha_x}}{\rm{(}}E_i,\,cos\theta{\rm{)}}=
\,\int d{\mathbf{r}}_i d{\mathbf{r}}_f
\Psi^{*{\rm{(}}-{\rm{)}}}_{{\mathbf{k}_f}}{\rm {(}}{\mathbf{r}}_f{\rm{)}}
I^{^*{\rm{(}}as{\rm{)}}}_{Aa;\,\alpha}({\mathbf{r}}_{Aa}{\rm{)}}
W^{{\rm{(}}as{\rm{)}}}_{ay;\,\alpha_x}{\rm{(}}{\mathbf{r}}_{ay}{\rm{)}}
 \Psi^{{\rm{(}}+{\rm{)}}}_{{\mathbf{k}_i}}{\rm {(}}{\mathbf{r}_i}{\rm{)}}.
\end{equation}

One notes that the expression (\ref{subeqSB23})  for $W^{{\rm{(}}as{\rm{)}}}_{ay;\,\alpha_x}{\rm{(}}{\mathbf{r}}_{ay}{\rm{)}}$  is  valid for $r_{ay}\gtrsim\, R_x$ and becomes identically   zero for $\eta_{ay}$= 0. In this case,   the  Fourier component of the $W^{{\rm{(}}as{\rm{)}}}_{x;\,\alpha_x}{\rm{(}}{\mathbf{r}}_{ay}{\rm{)}}$  function   in (\ref{subeqSB21}) is given only  by   the kinematic function $q_{ay}^{l_x}$ for $l_x\,>$ 0 and, so, the  Fourier integral becomes singular \cite{KMYa1988}. Therefore,   according    to \cite{KMYa1988},  for $\eta_{ay}$ = 0  this case  should be considered  specially by   putting $\eta_{ay}$ = 0 in the integrand of   Eq. (\ref{subeqSB21}) a priori. Then, for $\eta_{ay}$ = 0 one obtains  
\begin{equation}
\label{subeqSB26xas}
 W^{{\rm{(}}as{\rm{)}}}_{ay;\,\alpha_x}{\rm{(}}{\mathbf{r}}_{ay}{\rm{)}}=\,- \frac{C_{l_xj_x}}{{\rm{2}}\mu_{ay}}{\hat{l}}_x!! {\rm{(}}\kappa_{ay}r_{ay}{\rm{)}}^{-l_x}\delta{\rm{(}}r_{ay}{\rm{)}}r_{ay}^{-{\rm{2}}} Y_{l_x\nu_x}{\rm{(}}\hat{{\mathbf{r}}}_{ay}{\rm{)}},             
 \end{equation}where $r_{ay}$ is given by Eq. (\ref{subeqSB14}) and  ${\hat{l}}_x$= ${\rm{2}}l_x\,+\,{\rm{1}}$. This expression   corresponds to  the well-known zero-range approximation \cite{KMYa1988} and    can  be used jointly   with Eq. (\ref{subeqSB24}) in  the $  M^{{\rm{DWBA}}}_{{\rm{pole}}}{\rm{(}}E_i,\,cos\theta{\rm{)}}$  amplitude,  for example, for  the peripheral  $A$($d$, $n$)$B$ reaction.

 We now expand the $M^{{\rm{DWBA}}}_{{\rm{post;\, pole}}}{\rm{(}}E_i,\,cos\theta{\rm{)}}$ amplitude in  partial waves.  To this end, in (\ref{subeqSB26})  we use the  partial-waves expansions (A2) given in Appendix    and  the expansion 
$$
\frac{K_{l_{ay}\,+\,{\rm{3/2}}\,+\,\eta_{ay}}{\rm{(}}\kappa_{ay}r_{ay}{\rm{)}}}
{r_{ay}^{l_{ay}\,+\,\eta{ay}x\,+\,{\rm{3/2}}}}
\frac{K_{l_{Aa}\,+\,{\rm{1/2}}\,+\,\eta_{Aa}}{\rm{(}}\kappa_{Aa}r_{Aa}{\rm{)}}}
{r_{Aa}^{l_{Aa}\,+\,\eta_{Aa}\,+\,{\rm{1/2}}}}
$$
\begin{equation}
\label{subeqSB27}
=\,{\rm{4}}\pi\sum_{l\mu_l}{\cal{A}}_l{\rm{(}}r_i,\,r_f{\rm{)}}Y_{l\mu_l}{\rm{(}}\hat{{\mathbf r}}_i{\rm{)}}Y^*_{l\mu_l}{\rm{(}}\hat{{\mathbf r}}_f{\rm{)}}.
\end{equation}Here 
\begin{equation}
\label{subeqSB27}
{\cal{A}}_l{\rm{(}}r_i,\,r_f{\rm{)}}=\,\frac{\rm{1}}{\rm{2}}\int_{-{\rm{1}}}^{{\rm{1}}}\frac{K_{l_{ay}\,+\,{\rm{3/2}}\,+\,\eta_{ay}}{\rm{(}}\kappa_{ay}r_{ay}{\rm{)}}}
{r_{ay}^{l_{ay}\,+\,\eta_{ay}\,+\,{\rm{3/2}}}}
\frac{K_{l_{Aa}\,+\,{\rm{1/2}}\,+\,\eta_{Aa}}{\rm{(}}\kappa_{Aa}r_{Aa}{\rm{)}}}
{r_{Aa}^{l_{Aa}\,+\,\eta_{Aa}\,+\,{\rm{1/2}}}}P_l{\rm{(}}z{\rm{)}}dz,
\end{equation}where   $r_{ay}$= [($\bar{a}r_i$)$^{\rm{2}}$ + ($\bar{b}r_f$)$^{\rm{2}}$ - 2$\bar{a}\bar{b}r_ir_fz$]$^{\rm{1/2}}$,   $r_{Aa}$= [($\bar{c}r_i$)$^{\rm{2}}$ + ($\bar{d}r_f$)$^{\rm{2}}$ - 2$\bar{c}\bar{d}r_ir_fz$]$^{\rm{1/2}}$ and $z$\,=\,${\rm{(}}{\hat{{\mathbf{r}}}}_i{\hat{{\mathbf{r}}}}_f{\rm{)}}$. The integration over the angular variables ${\hat{\mathbf{r}}}_i $ and ${\hat{\mathbf{r}}}_f$ in Eq. (\ref{subeqSB26}) can easily be done by using Eqs. (A3) and (A4)   from Appendix. After some simple, but cumbersome algebra, one finds  that the pole amplitude $M^{{\rm{DWBA}}}_{{\rm{ pole}}}{\rm{(}}E_i,\,cos\theta{\rm{)}}$ in the system ${\it z}\Vert{\mathbf k}_i$ has the form 
$$
M^{{\rm{DWBA}}}_{{\rm{pole}}}{\rm{(}}E_i,\,cos\theta{\rm{)}}= -{\rm{8}}\sqrt{\frac{{\rm{2}}}{\pi}}
 \frac{{\rm{1}}}{\mu_{ay}}\frac{{\rm{1}}}{k_ik_f}\sum_{j_x\,\tau_x\,j_B\,\tau_B}\sum_{J,\,M
}\sum_{l_x\,l_B}{\rm{(-1)}}^{j_B-J_a+J}C_{ay;\,l_xj_x}C_{Aa;\,l_Bj_B}{\rm{(}}\hat{l}_x\hat{l}_B{\rm{)}}{\rm{(}}\hat{J}\hat{j}_B{\rm{)}}^{{\rm{1/2}}}
$$ 
 \begin{equation}
\label{subeqSB32}
\times i^{l_x+l_B}W{\rm{(}}l_xj_xl_Bj_B;J_aJ{\rm{)}}C_{j_B\tau_BJ_AM_A}^{J_BM_B} C_{j_x\tau_xJ_yM_y}^{J_xM_x}C_{JMj_B\tau_B}^{j_x\tau_x}\sum_{l_il_f}\,
e^{i\sigma_{l_i}+i\sigma_{l_f}}{\rm{(}}\hat{l}_i^{{\rm{2}}}\hat{l}_f{\rm{)}}^{{\rm{1/2}}} 
 \end{equation} 
 $$
\times {\rm{(}}\hat{l}_i^{{\rm{2}}}\hat{l}_f{\rm{)}}^{{\rm{1/2}}}C_{l_i\,{\rm{0}}l_f\,M}^{J\,M}A^{{\rm{pole}}}_{JMl_xl_Bl_il_f}Y_{l_fM}{\rm{(}}\theta,{\rm{0)}}, 
 $$ where the explicit form of  $A^{{\rm{pole}}}_{JMl_xl_Bl_il_f}{\rm{(}}k_i,\,k_f{\rm{)}}$ is  given by Eqs. (A7) -(A10) in  Appendix.
 
It should be noted that just neglecting  the dynamic recoil effect mentioned above, which is caused by using the pole approximation in the  matrix elements for the virtual decays $x\,\to\,y\,+\,a$ and     
 $B\,\to\,A\,+\,a$, results in the fact that the radial integral (A8) of the $ M^{{\rm{DWBA}}}_{{\rm{pole}}}{\rm{(}}E_i,\,cos\theta{\rm{)}}$ amplitude does not contain  the $V_{ya}$ and $V_{Aa}$ potentials  in contrast to that of the conventional DWBA with recoil effects \cite{Austern1964,LOLA}. That is the reason why  the $ M^{{\rm{DWBA}}}_{{\rm{ pole}}}{\rm{(}}E_i,\,cos\theta{\rm{)}}$ amplitude is parametrized  directly in the terms of the  ANCs  but not in  those of the spectroscopic factors as it occurs for the conventional DWBA \cite{Austern1964,LOLA}.

\vspace{1cm}  
  {\bf V.  THREE-PARTICLE COULOMB DYNAMICS OF THE TRANSFER MECHANISM AND THE GENERALIZED DWBA }
  \bigskip  

 \hspace{0.6cm}We now  consider  how to     accurately  take into account the contribution of the three-body Coulomb dynamics of  the transfer mechanism   to the $M^{{\rm{DWBA}}}_{{\rm{pole}}}{\rm{(}}E_i,\,cos\theta{\rm{)}}$ and $M^{{\rm{TBDWBA}}}{\rm{(}}E_i,\,cos\theta{\rm{)}}$ amplitudes by using Eqs. (\ref{subeqSB9}), (\ref{subeqSB10}) and (\ref{subeqSB32})  as well as  Eqs. (A7) and (A8) from  Appendix.   To this end, we should compare partial wave amplitudes $M^{{\rm{TB}}}_{l_i}{\rm{(}}E_i{\rm{)}}$ 
  and $M^{{\rm{DWBA}}}_{{\rm{pole}};\,l_i}$ for $l_i\,>>$ 1 determined from the corresponding expressions for    the $M^{{\rm(}s{\rm{)}}\,{\rm{TB}}}{\rm{(}}E_i,\,cos\theta{\rm{)}}$  and $M^{{\rm(}s{\rm{)}}\,{\rm{DWBA}}}_{{\rm{pole}}}{\rm{(}}E_i,\,cos\theta{\rm{)}}$ amplitudes \cite{Av86,Mukh10}.

According to \cite{Popov1964},   from  Eq. (\ref{subeqSB9}) and  (\ref{subeqSB10}), the   peripheral partial amplitudes at $l_i\,>>$ 1  and $l_f\,>>$ 1 can  be presented in the form as
 \begin{equation} 
M^{{\rm{TB}}}_{l_i\,l_f}{\rm{(}}E_i{\rm{)}}\,=\,{\tilde{\cal{R}}}^{{\rm{TB}}}{\rm{(}}E_i{\rm{)}}
M^{{\rm{DWBA}}}_{{\rm{pole}};\,l_i\,l_f}{\rm{(}}E_i{\rm{)}}.
\label{subeqSB12}
\end{equation}
  Here   $M^{{\rm{DWBA}}}_{{\rm{pole}};\,l_i\,l_f }{\rm{(}}E_i{\rm{)}}$ is the
 peripheral partial amplitude corresponding to  the pole approximation of the   DWBA amplitude  and ${\tilde{\cal{R}}^{{\rm{TB}}}}=\,{\tilde{N}}^{{\rm{TB}}}/{\tilde{N}}^{{\rm{DWBA}}}_{{\rm{pole}}}$, where  
          ${\tilde{N}}_{{\rm{pole}}}^{{\rm{DWBA}}}=\,N^{{\rm{DWBA}}}_{{\rm{pole}}}/\Gamma$, $\tilde{N}^{{\rm{TB}}}=\,N^{{\rm{TB}}}/\Gamma$
                    and $\Gamma\equiv\,\Gamma{\rm{(1}}\,-\,\eta_{ay}\,-\,\eta_{Aa}\,+\,i{\rm{(}}\eta_i\,+\,\eta_f{\rm{))}}$ is $\Gamma$-Euler's function. One notes that the CRF  $\tilde{\cal{R}}^{{\rm{TB}}}$ is  complex number and depends on the energy $E_i$, the binding energies $\varepsilon_{ay}$ and $\varepsilon _{Aa}$ as well as  the Coulomb parameters ($\eta_{ay}$, $\eta_{Aa}$, $\eta_i$ and $\eta_f$),  where  $\eta_i$ and $\eta_f$ are the Coulomb parameters for the entrance and exit channels, respectively. The expression for $M^{{\rm{TB}}}_{l_i\,l_f}{\rm{(}}E_i{\rm{)}}$ given by (\ref{subeqSB12}) can be considered as the peripheral partial amplitude of the generalized DWBA in which the contribution of    the three-body Coulomb dynamics of the main transfer mechanism is correctly taken into account.
As is seen from here, for $l_i\,>>$ 1 and $l_f\,>>$ 1 the asymptotics of      the pole approximation ($M^{{\rm{DWBA}}}_{{\rm{pole}};\,l_i\,l_f}{\rm{(}}E_i{\rm{)}}$)   of the DWBA and   the exact three-body partial amplitudes ($M^{{\rm{TB}}}_{l_i\,l_f}{\rm{(}}E_i{\rm{)}}$) have the same dependence on $l_i$ and $l_f$. Nevertheless,  they  differ only in their powers.

Therefore, if the main contribution to the    $M^{{\rm{TB}}}{\rm{(}}E_i,\,cos\theta{\rm{)}}$ amplitude   comes from the peripheral partial waves with $l_i\,>>$ 1  and $l_f\,>>$ 1, then the expression (\ref{subeqSB12}) makes it possible    to obtain the amplitude of  the generated DWBA. For this aim,    the expression ${\cal{B}}^{{\rm{pole}}}_{l_xl_Bl_il_f\lambda_{\rm{1}}\sigma_{\rm{1}}}{\rm{(}}k_i,k_f{\rm{)}}$, which enters   the pole approximation of the DWBA amplitude given by   Eq. (\ref{subeqSB32}) as well as Eqs. (A7) and (A8),  
  has to be renormalized by the   replacement  
\begin{equation}
 {\cal{B}}^{{\rm{pole}}}_{l_xl_Bl_il_f\lambda_{\rm{1}}\sigma_{\rm{1}}}{\rm{(}}k_i,k_f{\rm{)}}\,\longrightarrow\,\tilde{{\cal{B}}}^{{\rm{TB}}}_{l_xl_Bl_il_f\lambda_{\rm{1}}\sigma_{\rm{1}}}{\rm{(}}k_i,k_f{\rm{)}}=\,{\cal{N}}^{{\rm{TB}}}_{l_il_f}{\rm{(}}E_i{\rm{)}}{\cal{B}}^{{\rm{pole}}}_{l_xl_Bl_il_f\lambda_{\rm{1}}\sigma_{\rm{1}}}{\rm{(}}k_i,k_f{\rm{)}}.  
\label{subeqTBPWA1}
\end{equation}  Here
 \begin{equation}
{\cal{N}}^{{\rm{TB}}}_{l_il_f}{\rm{(}}E_i{\rm{)}}=\,\begin{cases}
{\rm{1}}, &{\rm{ for}} \, \,\, l_i,<\,L_{{\rm{0}}} \,\,{\rm{and}}\,\, l_f\,<\,L_{{\rm{0}}};\cr
 \tilde{\cal{R}}^{{\rm{TB}}}{\rm{(}}E_i{\rm{)}},\, &{\rm{for}}\,\, l_i\,\ge\,L_{{\rm{0}}}, \,\, l_f\,\ge\,L_{{\rm{0}}},\cr
\end{cases}
\label{subeqTBPWA2}
\end{equation} where $L_{{\rm{0}}}\sim k_iR_i^{{\rm{ch}}}$  (or $\sim k_fR_f^{{\rm{ch}}}$).

 From Eqs.  (\ref{subeqSB32}), (\ref{subeqTBPWA1}) and (\ref{subeqTBPWA2}), we can now  derive  the expression for the differential cross section for the generalized three-body DWBA.     It has   the form     as
$$
\frac{d\sigma}{d\Omega}=\frac{\mu_{Ax}\mu_{By}}{{\rm{(2}}\pi^{{\rm{2}}}{\rm{)}}^{{\rm{2}}}}\frac{k_f}{k_i}\frac{{\rm{1}}}{\hat{J}_A\hat{J}_x}\sum_{M_AM_xM_BM_y}\mid   M^{{\rm{(}}s{\rm{)}}{\rm{TB}}}{\rm{(}}E_i,\,\cos\theta)\mid^{\rm{2}} =\,\frac{{\rm{20}}}{\pi^{{\rm{3}}}}\frac{(\hbar c)^2}{E_iE_f}\left(\frac{\hbar}{\mu_{ay}c}\right)^2 \frac{k_f}{k_i} \frac{\hat{J}_B}{\hat{J}_A}
 $$ 
\begin{equation}
\label{FBCS1}
\times \sum_{j_x\,j_B}\sum_{J\,M}\mid
\sum_{l_x\,l_B}\sum_{l_i\,l_f}\,
\exp\{i[\sigma_{l_i}\,+\,\sigma_{l_f}\,+\,\frac{\pi}{{\rm{2}}}{\rm{(}}l_i\,+\,l_f\,+\,l_x\,+\,l_B)]\} 
 C_{ay;\,l_xj_x}C_{Aa;\,l_Bj_B}
\end{equation}
$$
\times {\rm{(}}\hat{l}_x\hat{l}_B{\rm{)}}{\rm{(}}\hat{l}_i^2\hat{l}_f{\rm{)}}^{{\rm{1/2}}} W{\rm{(}}l_xj_xl_Bj_B;J_aJ{\rm{)}}C_{l_i\,0l_f\,M}^{J\,M}
{\tilde{A}}^{{\rm{TB}}}_{JMl_xl_Bl_il_f}{\rm{(}}k_i,\,k_f{\rm{)}}Y_{l_f\,M}{\rm{(}}\theta,\,{\rm{0)}}
\mid^{{\rm{2}}},
$$
where the expression for ${\tilde{A}}^{{\rm{TB}}}_{JMl_xl_Bl_il_f}{\rm{(}}k_i,\,k_f{\rm{)}}$ is obtained from Eq. (A7) of Appendix by the substitution  of the ${\cal{B}}_{l_xl_Bl_il_f\lambda_{\rm{1}}\sigma_{\rm{1}}}{\rm{(}}k_i,k_f{\rm{)}}$  by  $\tilde{{\cal{B}}}^{{\rm{TB}}}_{l_xl_Bl_il_f\lambda_{\rm{1}}\sigma_{\rm{1}}}{\rm{(}}k_i,k_f{\rm{)}} $ given by  (\ref{subeqTBPWA1}). Herein,  the ANCs $C$'s , $\kappa_{ij}$ and $d\sigma/d\Omega$ are in fm$^{-{\rm{1/2}}}$, fm$^{-{\rm{1}}}$ and mb/sr, respectively. One notes that  Eqs. (\ref{FBCS1}) and (A8) given in  Appendix   contain the   cutoff parameters $R_i^{{\rm{ch}}}$ and  $ R_f^{{\rm{ch}}}$, which are determined by  only the free parameter $r_o$. Similar to how  it was done in Ref. \cite{KMYa1988},  the $r_o$ value can be  determined   by best fitting    the calculated     angular distributions to  the experimental ones  corresponding to the minimum of $\chi^{\rm{2}}$  at least in the angular region  of  the main peak. 

One notes that the expression (\ref{FBCS1}) can be considered as a  generalization of the dispersion theory proposed in   Ref. \cite{KMYa1988} for  the peripheral neutron transfer reactions  at above-barrier energies.    Nevertheless, the expression  (\ref{FBCS1}) can  also be applied for peripheral  strong sub-barrier charged particle transfer   reactions for which the dominant contribution comes to  rather low partial waves with $l_i$ (or $l_f$)= 0,1, 2,... In this case,  the  influence of the   three-body Coulomb  dynamics of the transfer mechanism  on the DCS is also  taken into account     via  the interference term between  the  low  and peripheral  partial  amplitudes   entering   Eq.   (\ref{FBCS1}) via Eq. (\ref{subeqTBPWA2}).

\vspace{1cm}  
  {\bf VI. RESULTS OF  APPLICATIONS TO THE SPECIFIC SUB- AND ABOVE-BARRIER REACTIONS }
  \bigskip  
  
   {\bf A. Asymptotic normalization coefficients}
  
\bigskip    
  
  In order      to verify  predictions of the asymptotic theory proposed in the present work and the influence of the three-body Coulomb dynamics of  the transfer mechanism  on the specific ANC values, we have calculated the differential cross sections of the proton and triton transfer reactions: ${\rm {^9Be(^{10}B,^9Be)^{10}B}}$  at the bombarding energy $E_{\rm{^{10}B}}$= 100 MeV
   \cite{Mukh2}; ${\rm {^{11}B^{12}C,^{11}B)^{12}C}}$ at $E_{\rm{^{12}C}}$= 87 MeV  \cite{LOLA};   ${\rm{^{16}O(^3He}}, d{\rm {)^{17}F}}$  at  $E_{\rm{^3He }}$= 29.75 MeV \cite{Mukh6}  and
 ${\rm{^{19}F(}}p,\,\alpha {\rm {)^{16}O}}$  at sub-barrier energies  $E_p$=  250; 350 and 450 keV \cite{HL1978,HAS1991} (denoted by EXP-1978 below) as well as $E_p$=  327; 387 and 486 keV \cite{Ivano2015} (denoted by EXP-2015 below). One notes that all these reactions are related to  the  ``non-dramatic'' case. 
 
  Calculations were performed     using   Eq. (\ref{FBCS1}) and the optical potentials in the initial and final  states, which were taken from Refs. \cite{Mukh2,LOLA,Mukh6,HAS1991}.  For these reactions the orbital ($l_x$ and $l_B$) and total ($j_x$ and $j_B$) angular momentums of the transfer particle (proton or triton) are  taken   equal to  $l_{{\rm{^{10}B}}}$=$l_{{\rm{^{12}C}}}$= 1,  $l_{{\rm{^{17}F_{(g.s.)}}}}$= 2 and $l_{{\rm{^3He}}}$=$l_{\alpha}$= 0, whereas $j_{{\rm{^{10}B}}}$=$j_{{\rm{^{12}C}}}$= 3/2,  $j_{{\rm{^{17}F^*}}}$= 5/2 and $j_{\alpha}$= $j_{{\rm{^3He}}}$= 1/2.
 
\begin{figure}
\begin{center}
\includegraphics[height=10cm, width=12cm]{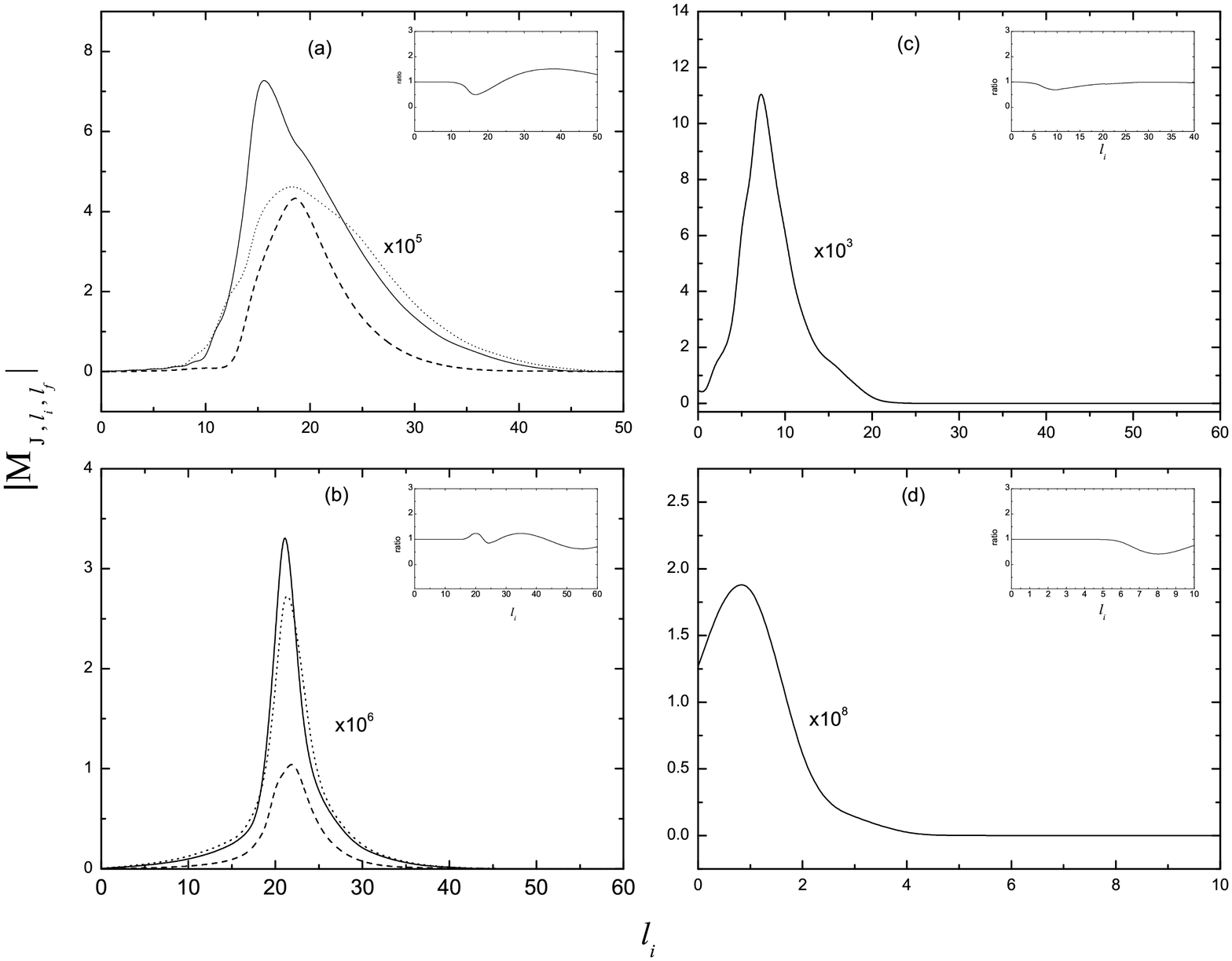}
 \end{center}
\caption{\label{fig3} The $l_i$ dependence of the modulus of the partial wave  amplitudes ($\mid M_{J\,l_i\,l_f}\mid$) for the ${\rm {^9Be(^{10}B,^9Be)^{10}B}}$ ($a$),   ${\rm {^{11}B(^{12}C,^{11}B)^{12}C}}$ ($b$),  $ {\rm{^{16}O(^3He}}, d{\rm {)^{17}F}}$(g.s.) ($c$) and  ${\rm{^{19}F(}}p,\,\alpha {\rm {)^{16}O}}$ ($d$)     reactions  at projectile energies of  $E_{\rm{^{10}B}}$= 100 MeV,  $E_{\rm{^{12}C}}$= 87 MeV,    $E_{\rm{^3He }}$= 29.75 MeV and  $E_p$=250 keV.  Here $l_i$ and  $l_f$ are  the relative orbital momenta in the entrance and exits channels of the considered reaction, respectively, and $J$ is the transferred angular momentum.  In ($a$) and ($b$),  the solid line is for $J$= 0 and $l_f=l_i $, the dashed line is for  $J$= 1 and $l_f=l_i\,+{\rm{1}} $  and  the dotted line is for  $J$= 2 and $l_f=l_i\,+{\rm{2}} $.  In ($c$),  the solid   line is  for  $J$= 2 ($l_f=l_i\,+{\rm{2}} $).  In ($d$), the solid line is for $J$= 0 ($l_f=l_i $).   The inserts are the  ratio of the   $\mid M_{J\,l_i\,l_f} \mid $   calculated with taking into account of the CRF of $\tilde{\cal{R}}^{{\rm{TB}}}{\rm{(}}E_i{\rm{)}}$ (see Eqs. (\ref{subeqTBPWA1}) and (\ref{subeqTBPWA2})) in the peripheral partial waves   to    that calculated with  $\tilde{\cal{R}}^{{\rm{TB}}}{\rm{(}}E_i{\rm{)}}$= 1 in the peripheral partial amplitudes.    } 
 \end{figure}

 The results of the  calculations of the  CRFs for the considered reactions, which  take into account the influence of the three-body Coulomb dynamics   on  the peripheral partial amplitudes, are listed in Table \ref{table2}. There, the  calculated values of  ${\tilde{N}}^{{\rm{DWBA}}}_{{\rm{post}}}$ correspond  to  the CRF  for the ``post'' form of DWBA  \cite{LOLA},  and   ${\tilde{\cal{R}}}^{{\rm{DWBA}}}_{{\rm{post}}}$ is determined by  the ratio of the  CRF  $\tilde{N}^{{\rm{TB}}}$ to that   
 ${\tilde{N}}^{{\rm{DWBA}}}_{{\rm{post}}}$ (${\tilde{\cal{R}}}^{{\rm{TB}}}_{{\rm{post}}}$=  
  $\tilde{N}^{{\rm{TB}}}/{\tilde{N}}^{{\rm{DWBA}}}_{{\rm{post}}}$), 
   where ${\tilde{N}}^{{\rm{DWBA}}}_{{\rm{post}}}$=$N^{{\rm{DWBA}}}_{{\rm{post}}}/\Gamma$  and   the explicit form of the CRF  $N^{{\rm{DWBA}}}_{{\rm{post}}}$ is determined  from  the expressions (14), (24)--(26) of Ref.   \cite{Mukh10}.  As is seen from  Table \ref{table2},   the differences between the calculated  CRFs  $ {\tilde{N}}^{{\rm{DWBA}}}_{{\rm{pole}}}$ and  $\tilde{N}^{{\rm{TB}}} $   are fairly  larger than that between the calculated  CRFs ${\tilde{N}}^{{\rm{DWBA}}}_{{\rm{post}}}$ and  $\tilde{N}^{{\rm{TB}}} $. This fact indicates that  the terms  $V_{yA}^C\,-\,V_f^C$ and $ \bigtriangleup V_f^CG_C\bigtriangleup V_i^C$ of the  transition operator, which enter the right-hand side of the amplitudes  Eqs. (\ref{subeqSB6}) and (\ref{subeqSB7}),  respectively, give a fairly large contribution to the peripheral partial amplitudes at $l_i\gtrsim k_iR_i^{\rm{ch}}>>$ 1  and $l_f\gtrsim k_fR_f^{\rm{ch}}>>$ 1 of the    
   $ M^{{\rm{TBDWBA}}}{\rm{(}}E_i,\,cos\theta{\rm{)}}$ amplitude  (\ref{subeqSB5}).  For an estimate of the influence of the CRFs on the calculated  peripheral partial amplitudes,
    we have  analyzed  the contribution of the  different  partial wave  amplitudes to  the  amplitude both for the   sub-barrier reactions and for the above-barrier one mentioned above.   Fig. \ref{fig3} shows  the $l_i$ dependence of the modulus of the  partial amplitudes, which are renormalized   on the product of the ANCs for the bound states of the nuclei in the entrance and exit channels.  As is seen from Figs. \ref{fig3}$a$ and \ref{fig3}$b$, the contribution to the amplitude of  the  ${\rm {^9Be(^{10}B,\,^9Be)^{10}B}}$ and  ${\rm {^{11}B(^{12}C,\,^{11}B)^{12}C}}$ reactions  from lower partial amplitudes with $l_i<$ 14 and $l_i<$ 15, respectively, is strongly suppressed  due to the strong absorption in the entrance and exit channels. Nevertheless,  for the transferred angular momentum  $J$= 0     the contributions of the three-body Coulomb effects    to  the modulus of the  partial amplitudes ($\mid M_{J\,l_i\,l_f} \mid $)     change from   55\%  to 7\%  for    the   ${\rm {^9Be(^{10}B,\,^9Be)^{10}B}}$  reaction at  $ l_i\ge$ 16  and  from 23\%  to 5\%  for  the  ${\rm {^{11}B(^{12}C,^{11}B)^{12}C}}$ one at  $ l_i\ge$ 21 (see the inset in Fig.\ref{fig3}).    It should be noted that the orbital angular momenta $l_i$ for these reactions are $l_i\sim k_iR_i^{{\rm{ch}}}\approx$ 16 and 21 for the channel radius $R_i^{{\rm{ch}}}\approx$ 5.3 and 5.6 fm, respectively. The analogous contribution  is found to be  about 20--30\%  for the  ${\rm{^{16}O(^3He}}, \,d{\rm {)^{17}F}}$(g.s.) reaction for which  $l_i\sim k_iR_i^{{\rm{ch}}}\approx$ 8   for the channel radius $R_i^{{\rm{ch}}}\approx$ 5 fm (see the inset in Fig.\ref{fig3}$c$).  For the  ${\rm{^{16}O(^3He}}, \,d{\rm {)^{17}F}}$(0.495 MeV) reaction  the influence   of the three-body Coulomb effects on  the peripheral partial amplitudes is extremely larger   as compared with that for  the  ${\rm{^{16}O(^3He}}, \,d{\rm {)^{17}F}}$(g.s.) reaction. For example, the  ratio of the   $\mid M_{J\,l_i\,l_f} \mid $   calculated with taking into account of the CRF of $\tilde{\cal{R}}^{{\rm{TB}}}{\rm{(}}E_i{\rm{)}}$ (see Eqs. (\ref{subeqTBPWA1}) and (\ref{subeqTBPWA2}))     to    that calculated without taking into account of  the CRF ($\tilde{\cal{R}}^{{\rm{TB}}}{\rm{(}}E_i{\rm{)}}$= 1)  in the peripheral partial amplitudes changes about from  1.3x10$^{{\rm -7}}$ to   2.2x10$^{{\rm -7}}$ for $l_i\ge$ 13.
   This is the result of the strong difference between  the ratio $\tilde{R}^{{\rm TB}}$ calculated for  the ground and fist exited states of the residual  ${\rm{^{17}F}}$ nucleus (see Table \ref{table2}).   In  Fig.  \ref{fig3}$d$, as an illustration,  the same $l_i$ dependence  is displayed     for the sub-barrier ${\rm{^{19}F(}}p,\,\alpha {\rm {)^{16}O}}$ reaction  at the energy  $E_p$=  0.250 MeV for which $l_i\sim k_iR_i^{{\rm{ch}}}\approx$ 1 corresponding to the channel radius $R_i^{{\rm{ch}}}\approx$ 5 fm. As is seen from Fig. \ref{fig3}$d$, the contribution of the peripheral partial wave to the reaction amplitude is strong suppressed, whereas the main contribution to the amplitude  comes to the low partial waves in the vicinity of $l_i\sim$ 1. The analogous  dependence occurs for other considered  energies $E_p$.  
   
    As is seen from here, the influence of the three-body Coulomb effects in the initial, intermediate and final states of  the considered above-barrier  reactions  on  the peripheral partial amplitudes of the reaction  amplitude   can not be ignored.  One notes that this influence is ignored  in the calculations  of
     the``post''-approximation and  the ``post'' form of the    DWBA performed in  \cite{Ar96} and \cite{Mukh2,LOLA}, respectively. In this connection,  it should be  noted    that this assertion is related also  to    the calculations of    the  dispersion peripheral  model  performed in  \cite{DDh1971}  for  the peripheral proton transfer reactions with   taking into account only  the   mechanism described  by the pole  diagram in  Fig. \ref{fig1}$a$.      Perhaps that is one of the possible reasons  why the NVC  (ANC) values  for the specific  virtual decay $B\to\,A\,+\,p$, derived   in \cite{DDh1971}      with and without  taking into account   the    Coulomb effects in the   vertices of the pole diagram  in Fig. \ref{fig1}$a$, differ strongly  from each other (see   Table in \cite{DDh1971}).
    
\begin{figure}
\begin{center}
\includegraphics[height=10cm, width=8cm]{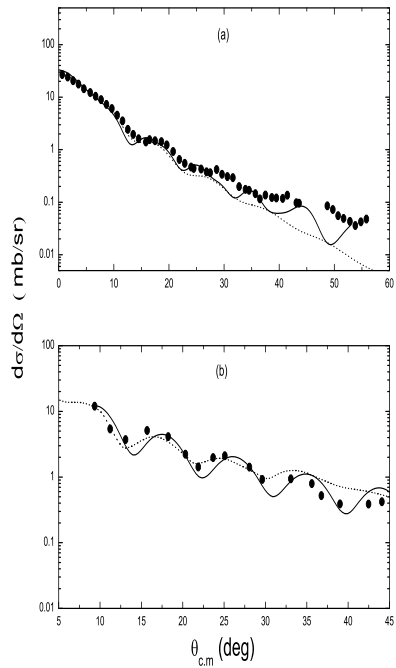}
 \end{center}
\caption{\label{fig4} The differential cross sections   for the ${\rm {^9Be(^{10}B}},\,{\rm{^9Be)^{10}B}}$ ($a$) and   ${\rm {^{11}B(^{12}C}},\,{\rm{^{11}B)^{12}C}}$ ($b$)  reactions at $E_{\rm{^{10}B}}$= 100 MeV and  $E_{\rm{^{12}C}}$= 87 MeV, respectively.  The  solid curves are the results of the present work, whereas the  dashed
lines are the results of  Refs. \cite{LOLA} and  \cite{Mukh2} derived in the conventional and modified DWBA, respectively. 
 The experimental data are taken from  Refs. \cite{Mukh2,LOLA}.}
\end{figure}

\begin{figure}
\begin{center}
\includegraphics[height=10cm, width=8cm]{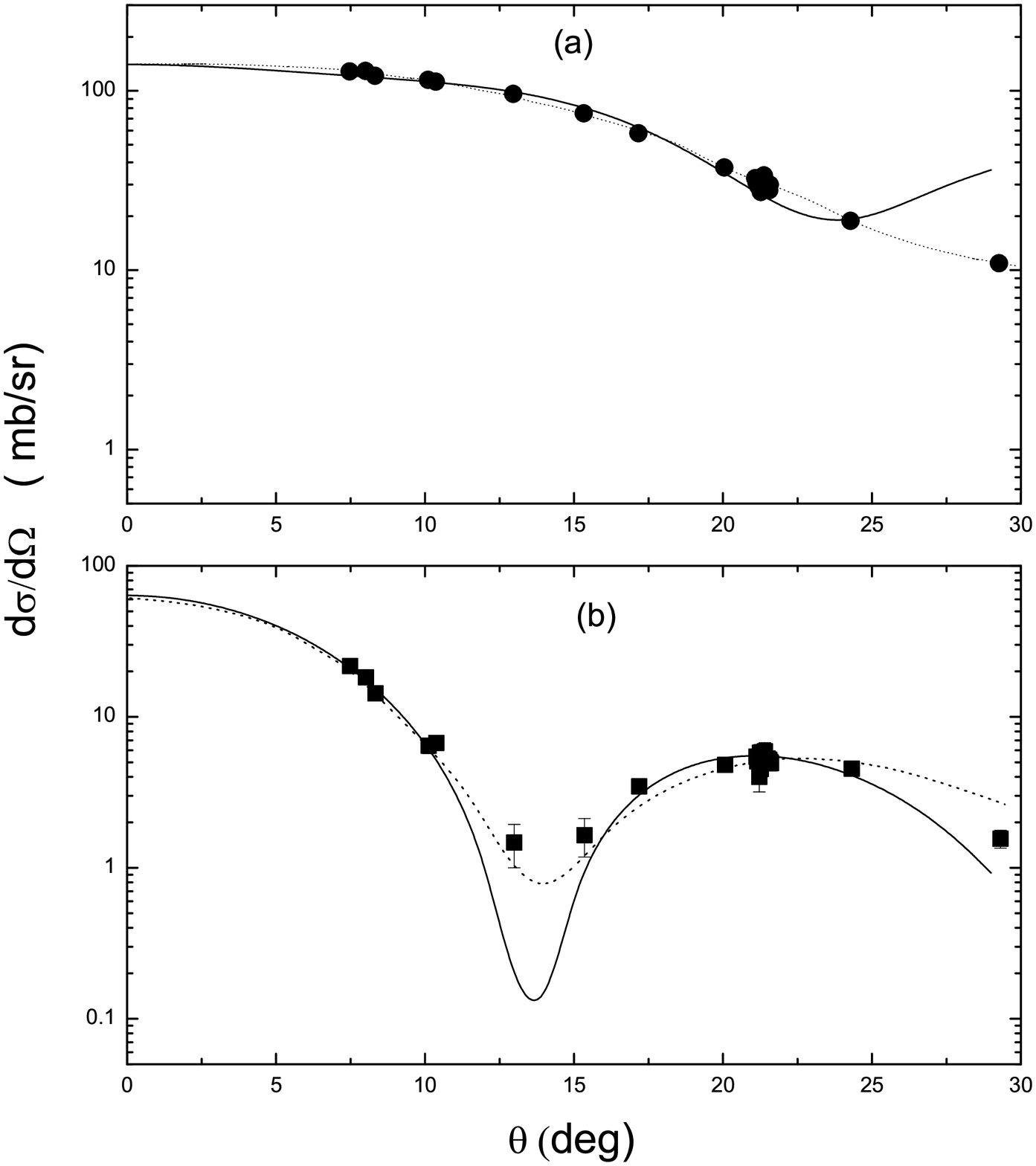}
 \end{center}
\caption{\label{fig5} The differential cross sections  for the  $ {\rm{^{16}O(^3He}}, d{\rm {)^{17}F}}$ reaction corresponding to   the ground ($a$) and first excited (0.429 MeV ($b$)) states of ${\rm{^{17}F}}$   at  $E_{\rm{^3He }}$= 29.75 MeV.  The  solid and dashed curves are the results of the present work and those of Ref.  \cite{Mukh6} derived  in  the ``post'' form of   the modified   DWBA.
 The experimental data are taken from  Refs. \cite{Mukh6}. }
\end{figure}

\begin{figure}
\begin{center}
\includegraphics[height=12cm, width=10cm]{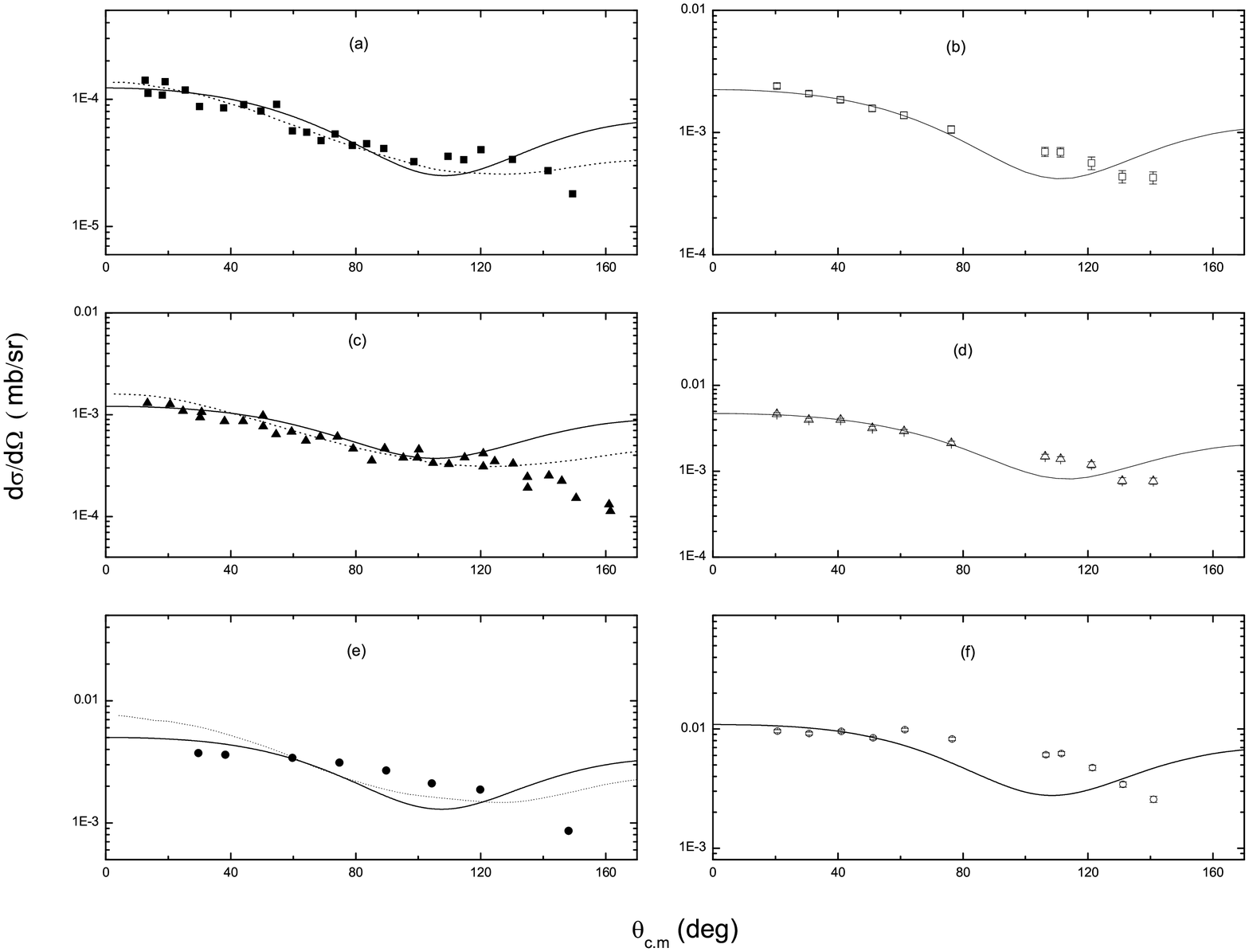}
 \end{center}
\caption{\label{fig6} The differential cross sections   for  the  ${\rm{^{19}F(}}p,\,\alpha {\rm {)^{16}O}}$ reaction  at  $E_p$=  450  ($a$),  350   ($b$) and 250 keV  ($c$) (the left side)  as well as  $E_p$= 327 ($d$),  387   ($e$) and 486 keV  ($f$) (the right side). The  solid curves are the results of the present work, whereas the  dashed lines are the results of Ref. \cite{HAS1991} derived in the zero-range of  the ``post''-approximation of DWBA.    The experimental data are taken from  Refs.  \cite{HL1978} (the EXP-1978:($a$), ($b$) and ($c$), see \cite{HAS1991} too)  and \cite{Ivano2015} (the EXP-2015:($d$), ($e$) and ($f$)).}
\end{figure}

    The results of the  calculations and  of the  comparison between the  differential cross sections obtained    in the present work (the solid curves), the  DWBA obtained in Refs. \cite{Mukh2,LOLA,HAS1991,Ivano2015}  by other authors (the dash curves) and experimental data are shown in Figs. \ref{fig4} -- \ref{fig6} and summarised in Table \ref{table3}. In the calculations we made use of the   two  different  optical potentials (the sets 1 and  2)   taken from   Refs. \cite{Mukh2,Mukh6}  and  the optical potentials taken from  \cite{LOLA,HAS1991}.   The results  of the present work      correspond to the standard value for the $r_{{\rm{0}}}$ parameter, which is taken  equal to 1.25 fm and  leads  also to   the  minimum     of $\chi^{\rm{2}}$     in  the  vicinity of  the main peak  of the angular distribution. It is seen that the angular distributions given by the asymptotic theory and the conventional  DWBA practically coincide and reproduce equally well the experimental ones.  The    ANCs (NVCs)  values  obtained in the present work, which are    presented in Table \ref{table3}, are found  by normalizing   the calculated    cross sections to  the  corresponding experimental ones at  the forward angles by  using  the ANCs $C^{\rm{2}}_{{\rm{^3He}}}$=4.20$\pm$0.32 fm$^{-1}$  \cite{YaBl2018} ($G^{\rm{2}}_{{\rm{^3He}}}$= 1.32$\pm$0.10 fm) for $d\,+p\to\,{\rm{^3He}}$  and $C^{\rm{2}}_{\alpha}$= 54.2$\pm$4.5 fm$^{-1}$\cite{Pl1973}  ($G^{\rm{2}}_{\alpha}$= 13.4$\pm$1.1 fm )  for  $t\,+p\to\,\alpha$. There, the   theoretical and experimental  uncertainties are  the result of   variation (up to $\pm$2.5\%) of  the cutoff $R_i^{{\rm{ch}}}$ and $ R_f^{{\rm{ch}}}$ (or $r_{{\rm{0}}}$) parameters    relative  to  the  standard $R_i^{{\rm{ch}}}$ and $ R_f^{{\rm{ch}}}$ ( $r_{{\rm{0}}}$= = 1.25 fm) values   and the experimental errors in $d^{\rm{exp}}/d\Omega$. 
    But,  the experimental uncertainties pointed out  in the  ANC (NVC) values for ${\rm{^{17}F}}\to\, {\rm{^{16}O }} \,+p$  and ${\rm{^{19}F}}\to\, {\rm{^{16}O }} \,+t$ correspond to  the mean  squared errors, which includes both the experimental errors in $d^{\rm{exp}}/d\Omega$ and  the above-mentioned uncertainty of the ANC (NVC)  for $d\,+p\to\,{\rm{^3He}}$ and  $t\,+p\to\,\alpha$, respectively. 
    
  As is seen from  Table \ref{table3}, the squared  ANC value for ${\rm{^9Be}} \,+p\to\, {\rm{^{10}B}}$ obtained in the present work differs  noticeably from that of   \cite{Mukh2} derived    from the analysis of  the same reaction performed within  the framework of the  ``post'' form of the modified DWBA.    Besides, as  can be  seen from Table \ref{table3}, the difference between   the squared  ANC values    obtained in the present work for    the set 1 and the set   2  of the optical potentials (the second and fifth lines)  does not exceed overall the experimental errors ($\Delta_{{\rm{exp}}}$= 7\%  \cite{Mukh2}) for the differential cross section, whereas such the difference  for the squared  ANC values derived  in \cite{Mukh2} (the third and sixth lines) exceeds noticeably $\Delta_{{\rm{exp}}}$      and is of  about  9\%. 
The  ANC  for ${\rm{^9Be}} \,+p\to\, {\rm{^{10}B}}$ recommended in the present work is presented in the seventh  and eighth  lines of Table  \ref{table3}, which has overall the uncertainty of  about  4\%. An  analogous situation occurs  when we  compare  the results   for   the ANC  for   ${\rm{^{16}O}}\,+p\to\,{\rm {^{17}F}}$(g.s)  and  ${\rm{^{16}O}}\,+p\to\,{\rm {^{17}F}}$(0.429 MeV)  between   the present work (the 19$^{\rm{th}}$  and   30$^{\rm{th}}$  lines), Ref.  \cite{Mukh6} ( the 20$^{\rm{th}}$   and 31$^{\rm{th}}$ lines) and  Ref. \cite{Ar96} ( the 21$^{\rm{th}}$ and  32$^{\rm{th}}$ lines).  
 Besides,  it is seen that the discrepancy   between  the ANC values of the present work and  Ref. \cite{Ar96} is larger  than that of \cite{Mukh6}. One notes once more  that the ``post''-approximation and the ``post'' form of the modified DWBA were used in  \cite{Ar96}   and \cite{Mukh6}, respectively. Nevertheless, the ANCs derived  in the present work  are in a good agreement within the uncertainties   with the results recommended in \cite{Ar2009}(see the 22$^{\rm{th}}$    and 33$^{\rm{th}}$ lines of  Table \ref{table3}).
 It is seen that  the asymptotic theory proposed in the present work provides better  accuracy for the ANC values   for $ {\rm{^9Be}} \,+p\to\,{\rm{^{10}B}}$   and ${\rm{^{16}O}}\, +\,p\to {\rm {^{17}F}}$  than that obtained  in Refs. \cite{Mukh2,Mukh6,Ar2009}.  
 
  The ANC values for 
  ${\rm{^{11}B}} \,+\,p\to\, {\rm{^{12}C}}$  and ${\rm{^{16}O}} \,+\,t\to \,{\rm{^{19}F}}$ obtained in the present work  are presented the eleventh and 34$^{\rm{th}}$  -- 43$^{\rm{th}}$ lines of Table   \ref{table3}, respectively. As is seen from there, the ANC  values   $ {\rm{^{16}O}} \,+\,t\to\,{\rm{^{19}F}}$   obtained  separately from the analysis of  the    experimental data taken from Refs.\cite{HL1978} (EXP-1978) and \cite{Ivano2015}(EXP-2015) differ from each other on the average  by  a factor of about  2.2.  This is the  result of the discrepancy between  the absolute  values of the experimental DCSs of  the EXP-1978 and the EXP-2015 measured independently  each other.   Because of a presence of  such the  discrepancy, we recommended decisive measurement of the experimental DCSs of the   ${\rm{^{19}F(}}p,\,\alpha {\rm {)^{16}O}}$  reaction  in the same energy region.  Nevertheless, one notes that the ANCs obtained separately  from the independent experimental data of the  EXP-1978 and EXP-2015
   at  the different projectile energies are stable, although the absolute  values of the corresponding experimental DCSs   depend strongly   on  the projectile energy (see Fig. \ref{fig6}). To best of our knowledge,  the ANC values for  $ {\rm{^{11}B}} \,+\,p \to\, {\rm{^{12}C}}$ and ${\rm{^{16}O}} \,+\,t\to\, {\rm{^{19}F}}$ presented in   Table \ref{table3} are obtained for the first time.
    
 \bigskip 
  
   {\bf B.  Astrophysical $S$ factors at stellar energies}
   
   \bigskip
   
 Here the weighted means of  the ANCs obtained for  ${\rm{^{16}O}}\, +\,p\to\, {\rm {^{17}F}}$(g.s.) and ${\rm{^{16}O}}\, +\,p\to\, {\rm ^{17}F}$(0.429 MeV),   ${\rm{^{11}B}}\,+p\to\,{\rm{^{12}C}}$(g.s.) and  $ {\rm{^9Be}} \,+p\to\,{\rm{^{10}B}}$(g.s.)    are used to calculate the   astrophysical $S$ factors for the   radiative capture   ${\rm{^{16}O(}}p,\,\gamma{\rm{)^{17}F}}$(g.s.), ${\rm{^{16}O(}}p,\,\gamma{\rm{)^{17}F}}$(0.429 \,MeV),  ${\rm{^9Be(}}p,\,\gamma{\rm{)^{10}B}}$(g.s.) and ${\rm{^{11}B(}}p,\,\gamma{\rm{)^{12}C}}$(g.s. )  reactions at stellar energies.  The calculations are performed  using  the modified two-body potential method \cite{Igam07} for the  direct radiative capture ${\rm{^{16}O(}}p,\,\gamma{\rm{)^{17}F}}$ reaction and the modified $R$-matrix  method only  for the direct component of astrophysical $S$ factors for the radiative capture  ${\rm{^9Be(}}p,\,\gamma{\rm{)^{10}B}}$(g.s.) and ${\rm{^{11}B(}}p,\,\gamma{\rm{)^{12}C}}$(g.s.) (see Ref. \cite{Ar2012} for example).

\begin{figure}
\begin{center}
\includegraphics[height=11cm, width=9cm]{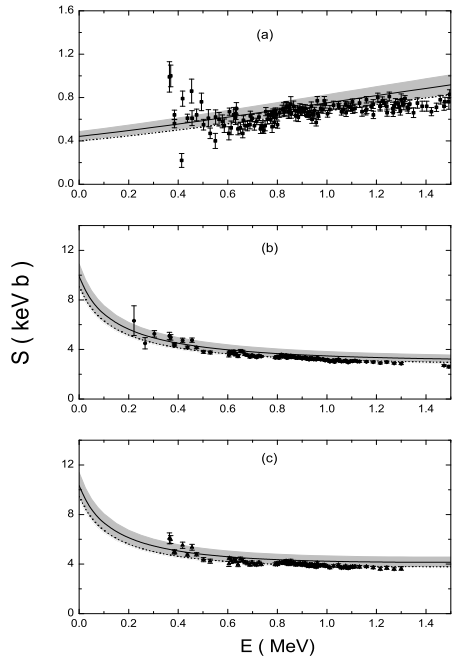}
\end{center}
\vspace{-2mm} \caption{\label{fig7}{\small The 
astrophysical $S$ factors for the direct radiative capture ${\rm {^{16}O(}}p,\gamma{\rm {)^{17}F}}$ reaction. The curves of  ($a$) and ($b$) correspond to  the ground and  first excited (0.429 MeV) states  of the residual ${\rm{^{17}F}}$ nucleus, respectively, whereas  that of  ($c$) corresponds to  their sum  
${\rm{^{17}F}}$ (g.s. + 0.429 MeV).
  The solid  and  the band  are  the results of
the present work, whereas   the  dashed  line is the result of   Ref. \cite{Ar2009}. The experimental data are  from \cite{Morl1997}.}}
\end{figure}

  Fig. \ref{fig7} shows the results of  comparison  between the  astrophysical $S$ factors  calculated  for the   radiative capture     ${\rm{^{16}O(}}p,\,\gamma{\rm{)^{17}F}}$  reaction and the experimental data  \cite{Morl1997}.  There, the  solid curves in  ($a$) and ($b$)   present the results obtained in the  present work   for    the ground and  first excited ($E^*$= 0.429 MeV) states   of  the residual ${\rm{^{17}F}}$ nucleus, respectively, whereas  the solid curve in ($c$) corresponds to  their sum  
   ${\rm{^{17}F}}$ (g.s. + 0.429 MeV). There, the bands are the corresponding uncertainties arising  because of the uncertainties of the ANCs. The  dashed lines in Fig. \ref{fig7} are taken from   Ref. \cite{Ar2009}.   As is seen from figure, the ANC values obtained in the present work for  ${\rm{^{16}O}}\, +\,p\to\, {\rm {^{17}F}}$(g.s.) and ${\rm{^{16}O}}\, +\,p\to\, {\rm{ ^{17}F}}$(0.429 MeV), firstly, reproduce well the  experimental data and, secondly,   allow extrapolation  of the astrophysical $S$ factors ($S$($E$)) at stellar energies. In a particular,  $S_{{\rm{g.s.}}}$($E$)= 0.44$\pm$0.04  and 0.45$\pm$0.05 keV$\cdot$b,  $S_{{\rm{exc.}}}$($E$)= 9.89$\pm$1.01  and 9.20$\pm$0.94 keV$\cdot$b and   $S_{{\rm{tot}}}$($E$)= 10.34$\pm$1.06  and 9.75$\pm$0.98 keV$\cdot$b are obtained  for $E$= 0 and 25 keV, respectively. One notes that our result   for $E$= 0  agrees     with that of  $S_{{\rm{tot}}}$(0)=9.45$\pm$0.4 keV$\cdot$b recommended in   \cite{Ar2009} as well as with the results of  10.2 and 11.0 keV$\cdot$b obtained  in \cite{BD1998} within the framework of the microscopic model for the effective  V2 and MN  potentials of the NN potential, respectively. 
 
  The results obtained  in the present work for  the direct component of  direct astrophysical $S$ factors ($S^{{\rm{DC}}}$($E$))  for  the ${\rm{^9Be(}}p,\,\gamma{\rm{)^{10}B}}$(g.s.) are  equal to $S^{{\rm{DC}}}$($E$)=0.173$\pm$0.0076 and 0.171$\pm$0.0075 keV$\cdot$b at  $E$=0 and 25 keV, respectively. Whereas,     for  the ${\rm{^{11}B(}}p,\,\gamma{\rm{)^{12}C}}$(g.s.) reaction they   are  equal to   $S^{{\rm{DC}}}$($E$)=0.190$\pm$0.008  and 0.187$\pm$0.008 keV$\cdot$b at  $E$=0 and 25 keV, respectively.      We note that  our result for  $S^{{\rm{DC}}}$(0) for  the ${\rm{^9Be(}}p,\,\gamma{\rm{)^{10}B}}$(g.s.) is in  a good agreement with that of \cite{Wulf1998} (within about 2$\sigma$)  and  larger  (by about  4.7$\sigma$)  than that recommended in \cite{Satt1999}.  This is a result of the discrepancy  between  the ANC values recommended in  the present work and Ref. \cite{Mukh2} (see Table \ref{table3}). 
   
 \bigskip  
  
  {\bf VII.  CONCLUSION }
  \bigskip

   \hspace{0.6cm}  Within the strong three-body  Schr\"{o}dinger formalism combined with the dispersion theory, a new asymptotic theory   is proposed for the  peripheral  sub- and above-barrier charged-particle transfer  $A$($x$, $y$)$B$ reaction, where  $x$=($y$ + $a$),  $B$=($A$ + $a$) and  $a$ is the transferred particle.  There,   the contribution of the three-body  ($A$, $a$ and $y$)  Coulomb dynamics of the  transfer  mechanism to the reaction amplitude is taken into account in a correct manner,   similar  to how       it is done  in the  dispersion theory.  Whereas,   an influence of  the distorted effects in the entrance and exit channels are kept in mind as it  is done in the conventional  DWBA.  The proposed asymptotic theory can be considered as a  generalization of the ''post''-approximation and  the ``post'' form of the conventional   DWBA in which the contributions of the three-body Coulomb effects in the initial, intermediate and final states  to the main pole mechanism is taken correctly  into account in all orders of the parameter of the perturbation theory over the Coulomb polarization potential  $V_{i,f}^C$.  
   
    The explicit form for  the differential cross section  (DCS) of the reaction under consideration  is obtained, which    is directly parametrized in the terms of the  product of  the squared  ANCs for  $y$ + $a\to\,x$  and    $A$ + $a\to\,B$ being  adequate to  the physics of the surface reaction. In the DCS, the  contributions both of  the rather low partial  waves and of the peripheral partial ones   are taken into account in a correct manner,  which   makes it possible to consider both the sub-barrier transfer reaction and the above-barrier one simultaneously.
  
  The  asymptotic  theory proposed in the present work  has been applied to the analysis of  the  experimental differential cross sections of the   specific above- and sub-barrier  peripheral   reactions corresponding to the proton and triton transfer mechanisms, respectively.  It is demonstrated   that     it       gives   an adequate description of both   angular distributions   in  the  corresponding main peaks of the angular distributions and    the absolute   values of the specific  ANCs (NVCs). The      ANCs   were  also applied to  calculations of the specific nuclear-astrophysical  radiative  proton capture reactions  and  new values of the   astrophysical $S$ factors extrapolated at stellar energies were obtained.

 \bigskip   
  {\bf ACKNOWLEDGEMENT}
   \bigskip

   This work has been support in part  by the Ministry of Innovations and Technologies   of the Republic of Uzbekistan (grant No. HE F2-14) and  by the Ministry of Education and Science of the Republic of Kazakhstan (grant No. AP05132062).  
   
    \bigskip
   
   {\bf APPENDIX: Formulae and expressions}
   
  \bigskip
  
  \hspace{0.6cm} Here we present the necessary formulae and expressions.
  
  The matrix element   $M_{Aa}{\rm{(}}{\mathbf{q}}_{Aa}{\rm{)}}$ of the virtual decay $B\,\to\,A\,+\,a$ is related to the  overlap function $I_{Aa}{\rm{(}}{\mathbf r}_{Aa}{\rm{)}}$ as \cite{Blok77}
$$
M_{Aa}{\rm{(}}{\mathbf{q}}_{Aa}{\rm{)}}\,=\,N_{Aa}^{{\rm{1/2}}}\int e^{-i{\mathbf{q}}_{Aa}{\mathbf{r}}_{Aa}}V_{Aa}{\rm{(}}{\mathbf{r}}_{Aa}{\rm{)}}I_{Aa}{\rm{(}}{\mathbf{ r}}_{Aa}{\rm{)}}d{\mathbf{ r}}_{Aa}
$$
$$
=-\,N_{Aa}^{{\rm{1/2}}}\Big(\frac{q_{Aa}^{{\rm{2}}}}{{\rm{2}}\mu_{Aa}}\,+\,\varepsilon_{Aa}\Big)
\int e^{-i{\mathbf{q}}_{Aa}{\mathbf{r}}_{Aa}}I_{Aa}{\rm{(}}{\mathbf{ r}}_{Aa}{\rm{)}}d{\mathbf{ r}}_{Aa} \eqno(A1)
$$
$$
=\,\sqrt{{\rm{4}}\pi}\sum_{l_B\mu_Bj_B\nu_B}C_{j_B\nu_BJ_AM_A}^{J_BM_B}C_{l_B\mu_B J_aM_a}^{j_B
 \nu_B}G_{Aa;\,l_Bj_B}{\rm{(}}q_{Aa}{\rm{)}}Y_{l_B\mu_B}(\hat{{\mathbf q}}_{Aa}{\rm{)}},
$$ 
  where  $G_{Aa;\,l_Bj_B}{\rm{(}}q_{Aa}{\rm{)}}$ is the vertex formfactor  for the virtual decay $B\,\to\,A\,+\,a$,  ${\mathbf{q}_{Aa}}$ is the relative momentum  of the $A$ and $a$ particles  and $G_{Aa;\,l_Bj_B}\equiv G_{Aa;\,l_Bj_B}{\rm{(}}i\kappa_{Aa}{\rm{)}}$, i.e., the NVC coincides with  the vertex formfactor 
   $G_{Aa; \,l_Bj_B}{\rm{(}}q_{Aa}{\rm{)}}$ when all the $B$, $a$ and $A$ particles are  on-shell  ($q_{Aa}=\,i\kappa_{Aa}$).
The same relations similar to Eq. (A1)  hold for the  matrix element   $M_{ay}{\rm{(}}{\mathbf{q}}_{ay}{\rm{)}}$ of the virtual decay $x\,\to\,y\,+\,a$ and the overlap function $I_{ay}{\rm{(}}{\mathbf r}_{ay}{\rm{)}}$.

The  partial-waves expansions for the distorted wave functions of  relative motion of the nuclei  in  the initial and exit states   of the reaction under consideration  have the form as \cite{Austern1964} 
 $$
 \Psi^{{\rm{(}}+{\rm{)}}}_{{\mathbf{k}_{i}}}{\rm {(}}{\mathbf{r}_{i}}{\rm{)}}=\,\frac{{\rm{4}}\pi}{k_ir_i}\sum_{l_i\mu_i}i^{l_i}e^{i\sigma_{l_i}}\Psi_{l_i}{\rm{(}}k_i;\,r_i{\rm{)}}Y_{l_i\mu_i}{\rm{(}}\hat{{\mathbf r}}_i{\rm{)}}Y^*_{l_i\mu_i}{\rm{(}}\hat{{\mathbf k}}_i{\rm{)}},
 $$
 $$
  \Psi^{*{\rm{(}}-{\rm{)}}}_{{\mathbf{k}_{i}}}{\rm {(}}{\mathbf{r}_{i}}{\rm{)}}=\,\frac{{\rm{4}}\pi}{k_fr_f}\sum_{l_f\mu_f}i^{-\,l_f}e^{i\sigma_{l_f}}\Psi_{l_f}{\rm{(}}k_f;\,r_f{\rm{)}}Y_{l_f\mu_f}{\rm{(}}\hat{{\mathbf r}}_f{\rm{)}}Y^*_{l_f\mu_f}{\rm{(}}\hat{{\mathbf r}}_f{\rm{)}},\eqno(A2)
 $$  
where $\Psi_{l}{\rm{(}}k;\,r{\rm{)}}$ is the partial wave functions in the initial state or the final one. 

 The expansions of the $r_{ay}^{l_x}Y_{l_x\sigma_x}{\rm{(}}{\hat{\mathbf{r}}}_{ay}{\rm{)}}$ and $r_{Aa}^{l_B}Y_{l_B\sigma_B}^*{\rm{(}}{\hat{\mathbf{r}}}_{Aa}{\rm{)}}$ functions on the bipolar harmonics of the $l_x$ rank  and the $l_B$ one have the forms as 
$$
r_{ay}^{l_x}Y_{l_x\sigma_x}{\rm{(}}{\hat{\mathbf{r}}}_{ay}{\rm{)}}=\,\sqrt{{\rm{4}}\pi}\sum_{\lambda_1+\lambda_2=\,l_x}\,\,\sum_{\tilde{ \mu}_{\lambda_1}\tilde{\mu}_{\lambda_2}}
\Big(\frac{\hat{l}_x!}{\hat{\lambda}_1!{\hat{\lambda}}_2!}\Big)^{{\rm{1/2}}}
\Big(\frac{\mu_{Ax}}{m_a}r_i\Big)^{\lambda_1}
\Big(-\frac{\mu_{Ax}}{\mu_{Aa}}r_f\Big)^{\lambda_2}
$$
$$
\times C_{\lambda_1\tilde{\mu}_{\lambda_1}\,\lambda_2\tilde{\mu}_{\lambda_2}}^{l_x\mu_x}Y_{\lambda_1\tilde{\mu}_{\lambda_1}}{\rm{(}}{\hat{\mathbf{r}}}_i{\rm{)}}Y_{\lambda_2\tilde{\mu}_{\lambda_2}}{\rm{(}}{\hat{\mathbf{r}}}_f{\rm{)}}\eqno(A3)
$$ and
$$
r_{Aa}^{l_B}Y_{l_B\sigma_B}^*{\rm{(}}{\hat{\mathbf{r}}}_{Aa}{\rm{)}}=\,\sqrt{{\rm{4}}\pi}\sum_{\sigma_1+\sigma_2=\,l_B}\,\,\sum_{\tilde{ \mu}_{\sigma_1}\tilde{\mu}_{\sigma_2}}
\Big(\frac{\hat{l}_B!}{\hat{\sigma}_1!{\hat{\sigma}}_2!}\Big)^{{\rm{1/2}}}
\Big(-\frac{\mu_{By}}{\mu_{ay}}r_i\Big)^{\sigma_1}
\Big(\frac{\mu_{By}}{m_a}r_f\Big)^{\sigma_2}
$$
$$
\times C_{\sigma_1\tilde{\mu}_{\sigma_1}\,\sigma_2\tilde{\mu}_{\sigma_2}}^{l_B\mu_B}Y^*_{\sigma_1\tilde{\mu}_{\sigma_1}}{\rm{(}}{\hat{\mathbf{r}}}_i{\rm{)}}Y^*_{\sigma_2\tilde{\mu}_{\sigma_2}}{\rm{(}}{\hat{\mathbf{r}}}_f{\rm{)}}.
\eqno(A4)
$$  Eqs. (A3) and (A4) can be derived from (\ref{subeqSB14}) and    
 $$
\int d{\hat{\mathbf{r}}}_iY^*_{\sigma_1\tilde{\mu}_{\sigma_1}}{\rm{(}}{\hat{\mathbf{r}}}_i{\rm{)}}Y_{l\mu_l}{\rm{(}}\hat{{\mathbf r}}_i{\rm{)}}Y_{l_i\mu_{l_i}}{\rm{(}}\hat{{\mathbf r}}_i{\rm{)}}Y_{\lambda_1\tilde{\mu}_{\lambda_1}}{\rm{(}}{\hat{\mathbf{r}}}_i{\rm{)}}=\, {\rm{(-1)}}^{\tilde{\mu}_l}\sum_{I{\tilde{\mu}}_I}
\left(\frac{\hat{l_i}\hat{\lambda}_1\hat{l}\hat{\sigma}_1}{{\rm{(4}}\pi{\rm{)}}^{\rm{2}}\hat{I}\hat{I}}\right)^{{\rm{1/2}}}
$$
$$
\times C_{l_i\,{\rm{0}}\,\lambda_{{\rm{1}}}\,{\rm{0}}}^{I\,{\rm{0}}}
C_{l\,{\rm{0}}\,\sigma_{{\rm{1}}}\,{\rm{0}}}^{I\,{\rm{0}}}
C_{l_i\,\tilde{\mu}_{l_i}\,\lambda_{{\rm{1}}}\,\tilde{\mu}_{\lambda_{{\rm{1}}}}}^{I\,\tilde{\mu}_I}
C_{l\,-\tilde{\mu}_l\,\sigma_{{\rm{1}}}\,\tilde{\mu}_{\sigma_{\rm{1}}}}^{I\,\tilde{\mu}_I}, \eqno(A5)
$$ 
 $$
\int d{\hat{\mathbf{r}}}_fY^*_{l\mu_l}{\rm{(}}\hat{{\mathbf r}}_f{\rm{)}}Y^*_{\sigma_{\rm{2}}\tilde{\mu}_{\sigma_{\rm{2}}}}{\rm{(}}{\hat{\mathbf{r}}}_f{\rm{)}}Y_{l_f\mu_{l_f}}{\rm{(}}\hat{{\mathbf r}}_f{\rm{)}}Y_{\lambda_{{\rm{2}}}\tilde{\mu}_{\lambda_{{\rm{2}}}}}{\rm{(}}{\hat{\mathbf{r}}}_f{\rm{)}}=\, \sum_{L{\tilde{\mu}}_L}
\left(\frac{\hat{l_f}\hat{\lambda}_{{\rm{2}}}\hat{l}\hat{\sigma}_{{\rm{2}}}}{{\rm{(4}}\pi{\rm{)}}^{\rm{2}}\hat{L}\hat{L}}\right)^{{\rm{1/2}}}
$$
$$
\times C_{l_f\,0\,\lambda_{{\rm{2}}}\,{\rm{0}}}^{L\,{\rm{0}}}C_{l\,{\rm{0}}\,\sigma_{{\rm{2}}}\,{\rm{0}}}^{L\,{\rm{0}}}C_{l_f\,\tilde{\mu}_{l_f}\,\lambda_{{\rm{2}}}\,\tilde{\mu}_{\lambda_{{\rm{2}}}}}^{L\,\tilde{\mu}_L}
C_{l\,\tilde{\mu}_{l}\,\sigma_{{\rm{2}}}\,\tilde{\mu}_{\sigma_{{\rm{2}}}}}^{L\,\tilde{\mu}_L}.
\eqno(A6)
$$ 

The explicit form of  $A^{{\rm{pole}}}_{JMl_xl_Bl_il_f}{\rm{(}}k_i,\,k_f{\rm{)}}$ entering Eq. (\ref{subeqSB32}) is given by 
 $$
A^{{\rm{pole}}}_{JMl_xl_Bl_il_f}{\rm{(}}k_i,\,k_f{\rm{)}}=\sum_{\sigma_{{\rm{1}}}+\sigma_{{\rm{2}}}=l_B\,\,\,} \sum_{\lambda_{{\rm{1}}}+\lambda_{{\rm{2}}}=l_x}\sum_{lIL}\hat{l}
\left(
\begin{array}{cc}
2l_{x}\\
2\lambda_1
\end{array}
\right)^{{\rm{1/2}}}
$$
$$
\times\left(
\begin{array}{cc}
{\rm{2}}l_{B}\\
{\rm{2}}\sigma_{{\rm{1}}}
\end{array}
\right)^{{\rm{1/2}}}
{\bar{a}}^{\lambda_{{\rm{1}}}}
{\bar{b}}^{\lambda_{{\rm{2}}}}
{\bar{c}}^{\sigma_{{\rm{1}}}}{\bar{d}}^{\sigma_{{\rm{2}}}}
C_{l\,{\rm{0}}\sigma_{{\rm{1}}}\,{\rm{0}}}^{I\,{\rm{0}}}C_{l_i\,{\rm{0}}\lambda_1\,{\rm{0}}}^{I\,{\rm{0}}}
C_{l\,{\rm{0}}\,\sigma_{{\rm{ 2}}}\,{\rm{0}}}^{L\,{\rm{0}}}C_{l_f\,{\rm{0}}\lambda_{{\rm{2}}}\,{\rm{0}}}^{L\,{\rm{0}}}
\eqno(A7)
$$
$$
\times W{\rm{(L}}\sigma_{{\rm{2}}}I\sigma_{{\rm{1}}};\,ll_B{\rm{)}}X{\rm{(}}\lambda_{{\rm{1}}}\lambda_{{\rm{2}}}l_x;\,l_il_fJ;\,ILl_B{\rm{)}}{\cal{B}}^{{\rm{pole}}}_{l_xl_Bl_il_f\lambda_{{\rm{1}}}\sigma_{{\rm{1}}}}{\rm{(k}}_i,\,k_f{\rm{)}},
$$
 $$
{\cal{B}}^{{\rm{pole}}}_{l_xl_Bl_il_f\lambda_{\rm{1}}\sigma_{\rm{1}}}{\rm{(}}k_i,\,k_f{\rm{)}}={\rm{(}}\eta_{ay}/4^{\eta_{ay}+\eta_{Aa}}{\rm{)}}{\rm{(}}\kappa_{Aa}/{\rm{2}}{\rm{)}}^{l_B}{\rm{(}}\kappa_{ay}/{\rm{2}}{\rm{)}}^{l_x}\kappa_{Aa}\kappa_{ay}^{\rm{3}}
$$
$$
\times \int_{R_i^{{\rm{ch}}}}^{\infty}dr_ir_i^{\lambda_{\rm{1}}\,+\,\sigma_{\rm{1}}\,+\,{\rm{1}}}\Psi_{l_i}{\rm{(r}}_i;k_i{\rm{)}}
\int_{R_f^{{\rm{ch}}}}^{\infty}dr_fr_f^{\lambda_{\rm{2}}\,+\,\sigma_{\rm{2}}\,+\,{\rm{1}}}\Psi_{l_f}{\rm{(r}}_f;\,k_f{\rm{)}}{\tilde{\cal{A}}}_{l_Bl_xl}{\rm{(}}r_i,\,r_f{\rm{)}}, \eqno(A8)
$$
 $$
{\tilde{\cal{A}}}_{l_Bl_xl}{\rm{(}}r_i,\,r_f{\rm{)}}=\,\frac{{\rm{1}}}{{\rm{2}}}\int_{-{\rm{1}}}^{{\rm{1}}}dz P_l{\rm{(}}z{\rm{)}}F_{l_B}{\rm{(}}r_{Aa};\kappa_B,\eta_{Aa}-{\rm{1)}}
F_{l_x}{\rm{(}}r_{ay};\kappa_{ay},\eta_{ay}{\rm{)}},\eqno(A9)
$$
 $$
 F_{l}(r;\kappa,\eta) =\frac{\pi^{1/2}}{\Gamma(l+\eta+2)}\int_{1}^{\infty}dte^{-\kappa rt}  (t^2-1)^{l+\eta+1},
 \eqno(A10)
$$
where $W{\rm{(}}l_{{\rm{1}}}j_{{\rm{1}}}l_{{\rm{2}}}j_{{\rm{2}}};j_{{\rm{3}}}j_{{\rm{4}}}{\rm{)}}$  and $X{\rm{(}}\lambda_{{\rm{1}}}\lambda_{{\rm{2}}}l_x;\,l_il_fJ;\,ILl_B{\rm{)}}$ are Rakah and Fano coefficients \cite{Varsh}, respectively; $R_i^{{\rm{ch}}}=\,R_x\,+\,R_A$ ($R_f^{{\rm{ch}}}=\,R_y\,+\,R_B$) is  the cutoff radius in the entrance (exit) channel;  $\left(m\atop n\right)$ is the binomial coefficient  and $\hat{j}$= ${\rm{2}}j\,+\,{\rm{1}}$.

\newpage

\newpage

 \newpage
 
\newpage
 
\newpage

 \newpage  
 
 \newpage

\newpage

\newpage

{\selectlanguage{english}
\begin{table}[t]
\begin{center}
 \caption{\label{table1}{\small The specific reactions and the corresponding to them vertices  described by the triangle diagram  Fig. \ref{fig2}$c$,   the positions of  singularities $i\kappa$ and $i\kappa_i$ ($i\bar{\kappa}_i$) in $q_{Aa}$($q_{ay}$) as well as  $\xi$ and $\xi_i$ ($\bar{\xi}_i$) in  the  $\cos\theta$-plane of the reaction amplitude, where $\kappa$ is related either  to the vertex B $\to$ A+$a$ ( $\kappa$=  $\kappa_{Aa}$)  or to the vertex x $\to$ y+$a$ ($\kappa$= $\kappa_{ya}$) }.}
 
 \vspace{2mm}
{\footnotesize
\begin{tabular}{lllllllll} \hline
\multicolumn{1}{c}{}
&\multicolumn{1}{c}{}
&\multicolumn{1}{c}{ The vertex}
&\multicolumn{1}{c}{}
&\multicolumn{1}{c}{}
&\multicolumn{1}{c}{}
&\multicolumn{1}{c}{}
&\multicolumn{1}{c}{}
&\multicolumn{1}{c}{}\\
\multicolumn{1}{c}{Reaction}
&\multicolumn{1}{c}{$E_x^{{\rm{lab}}}$}
&\multicolumn{1}{c}{$B\,\to\,A\,+\,a$}
&\multicolumn{1}{c}{ $\xi$}
&\multicolumn{1}{c}{$b_i$ }
&\multicolumn{1}{c}{$c_i$}
&\multicolumn{1}{c}{$d_i$}
&\multicolumn{1}{c}{$\kappa_i$($\bar{\kappa}_i$), }
 &\multicolumn{1}{c}{$\xi_i$} \\
\multicolumn{1}{c}{$A$($x$,\,$y$)$B$}
&\multicolumn{1}{c}{MeV}
&\multicolumn{1}{c}{($x\,\to\,y\,+\,a$)}
&\multicolumn{1}{c}{($\kappa$, fm$^{{\rm{-1}}}$)}
&\multicolumn{1}{c}{($e_i$) }
&\multicolumn{1}{c}{($f_i$)}
&\multicolumn{1}{c}{($g_i$)}
&\multicolumn{1}{c}{fm$^{\rm{{-1}}}$}
 &\multicolumn{1}{c}{($\bar{\xi}_i$)} \\
 \hline\\
\multicolumn{1}{c}{${\rm{^9Be}}$(${\rm{^{10}B}}$,\,${\rm{^9Be}}$)${\rm{^{10}B}}$}  
&\multicolumn{1}{c}{100} 
 &\multicolumn{1}{c}{${\rm{^{10}B}}\to\,{\rm{^9Be}}\,+\,p$}
&\multicolumn{1}{c}{1.020(0.534)}
&${\rm{^8Be}}$
&$\,\,\,\,d$
&$\,\,\,\,n$
&0.940&1.064\\
 \multicolumn{1}{c}{  } 
& \multicolumn{1}{c}{  }   
& \multicolumn{1}{c}{  } 
 &\multicolumn{1}{c}{}
 &${\rm{^6Li}}$
 &${\rm{^4{He}}}$
 &$\,\,\,\,t$
 &2.024
 &1.479\\
 \multicolumn{1}{c}{  } 
& \multicolumn{1}{c}{  }   
& \multicolumn{1}{c}{  } 
 &\multicolumn{1}{c}{}
 &$\,\,\,\,n$
 &${\rm {^9{Be}}}$
 &${\rm {^8{Be}}}$
 &0.802
 &4.169\\
 \multicolumn{1}{c}{${\rm{^{11}B}}$(${\rm{^{12}C}}$,\,${\rm{^{11}B}}$)${\rm{^{12}C}}$}
&\multicolumn{1}{c}{87}
&\multicolumn{1}{c}{${\rm{^{12}C}}\to\,{\rm{^{11}B}}\,+\,p$}
&\multicolumn{1}{c}{1.037(0.840)}
&${\rm{^{10}B}}$
&$\,\,\,\,d$
&$\,\,\,\,n$
&2.131
&1.264\\
\multicolumn{1}{c}{  } 
& \multicolumn{1}{c}{  }   
& \multicolumn{1}{c}{  } 
 &\multicolumn{1}{c}{}
 &${\rm{^8Be}}$
 &${\rm{^4{He}}}$
 &$\,\,\,\,t$
 &2.059
 &1.384\\
 \multicolumn{1}{c}{  } 
& \multicolumn{1}{c}{  }   
& \multicolumn{1}{c}{  } 
 &\multicolumn{1}{c}{}
  &$\,\,\,\,n$
  &${\rm{^{11}C}}$
  &${\rm{^{10}B}}$
   &1.618
    &16.020\\
     \multicolumn{1}{c}{${\rm{^{16}O}}$(${\rm{^3He}}$,\,$d$)${\rm{^{17}F}}$(g.s.)}
&\multicolumn{1}{c}{29.7}
&\multicolumn{1}{c}{${\rm{^{17}F}}\to\,{\rm{^{16}O}}\,+\,p$}
&\multicolumn{1}{c}{1.065(0.165)}
&${\rm{^{14}N}}$
&${\rm{^3He}}$
&$\,\,\,\,d$
&2.696
&3.253\\
\multicolumn{1}{c}{  } 
& \multicolumn{1}{c}{  }   
& \multicolumn{1}{c}{  } 
 &\multicolumn{1}{c}{}
 &${\rm{^{13}N}}$
&${\rm{^4{He}}}$
&$\,\,\,\,t$
&2.645
&3.508\\
\multicolumn{1}{c}{  } 
& \multicolumn{1}{c}{  }   
& \multicolumn{1}{c}{  } 
 &\multicolumn{1}{c}{}
 &$\,\,\,\,p$
 &${\rm{^{16}O}}$
 &${\rm{^{15}N}}$
  &0.905
   &49.551\\
   \multicolumn{1}{c}{} 
& \multicolumn{1}{c}{}
 &\multicolumn{1}{c}{(${\rm{^3He}}\to\,d\,+\,p$)}
 & \multicolumn{1}{c}{ 1.065( 0.420)}
 &\,\,\,\,($p$)
 &\,\,\,\,($d$)
 &\,\,\,\,($n$)
 &(0.652)
 &(1.562)\\
   \multicolumn{1}{c}{${\rm{^{19}F}}$($p$,\,$\alpha$)${\rm{^{16}O}}$}
&\multicolumn{1}{c}{0.250}
&\multicolumn{1}{c}{${\rm{^{19}F}}\to\,{\rm{^{16}O}}\,+\,t$}
&\multicolumn{1}{c}{13.648(1.194)}
&${\rm{^{15}N}}$
&$\alpha$
&$\,\,\,\,d$
&1.522
&19.720\\
 \multicolumn{1}{c}{ }
&\multicolumn{1}{c}{0.350}
&\multicolumn{1}{c}{ }
&\multicolumn{1}{c}{11.544(1.194)}
 & 
& & 
&1.522
&16.647\\
 \multicolumn{1}{c}{ }
&\multicolumn{1}{c}{0.450}
&\multicolumn{1}{c}{ }
&\multicolumn{1}{c}{10.190(1.194)}
 & 
& & 
&1.522
&14.665\\
\hline
 \end{tabular}}
\end{center}
\end{table}}

\newpage

{\selectlanguage{english}
 \begin{table}[t]
\begin{center}
 \caption{\label{table2}{\small  Reaction $A$($x$, $y$)$B$, incident energy $E_x$,  values of CRFs  $ {\tilde{N}}^{{\rm{DWBA}}}_{{\rm{pole}}}$ and   $ {\tilde{N}}^{{\rm{DWBA}}}_{{\rm{post}}}$  as well as  $\tilde{N}^{{\rm{TB}}} $ in the  pole-approximation and  the ''post'' form of DWBA as well as  the three-body model, respectively,  and quantities $ {\tilde{\cal{R}}^{{\rm{TB}}}_{{\rm{post}}}}$= $\tilde{N}^{{\rm{TB}}}/{\tilde{N}}^{{\rm{DWBA}}}_{{\rm{post}}}$, $ {\tilde{\cal{R}}^{{\rm{TB }}}}$=   $\tilde{N}^{{\rm{TB}}}/{\tilde{N}}^{{\rm{DWBA}}}_{{\rm{pole}}}$ and $ {\tilde{\cal{R}}^{{\rm{DWBA}}}_{{\rm{post}}}}=\,{\tilde{\cal{R}}^{{\rm{TB }}}}/ {\tilde{\cal{R}}^{{\rm{TB}}}_{{\rm{post}}}}$. }}
 
 \vspace{5mm}
{\footnotesize
\begin{tabular}{lllll}
 \hline
 \multicolumn{1}{c}{} &\multicolumn{1}{c}{}&\multicolumn{1}{c}{}&\multicolumn{1}{c}{}&\multicolumn{1}{c}{}\\
\multicolumn{1}{c}{$A$($x$, $y$)$B$}
 &\multicolumn{1}{c}{\,\,\,  $E_x$, MeV}
  &\multicolumn{1}{c}{$ {\tilde{N}}^{{\rm{DWBA}}}_{{\rm{pole}}}$ ($ {\tilde{N}}^{{\rm{DWBA}}}_{{\rm{post}}}$)}
&\multicolumn{1}{c}{\,\,\,$\tilde{N}^{{\rm{TB}}} $  }
&\multicolumn{1}{c}{$ {\tilde{\cal{R}}}^{{\rm{TB}}}$ ($ {\tilde{\cal{R}}}^{{\rm{TB}}}_{{\rm{post}}}$) }\\
\multicolumn{1}{c}{ }
 &\multicolumn{1}{c}{}
  &\multicolumn{1}{c}{ }
&\multicolumn{1}{c}{   }
&\multicolumn{1}{c}{  [${\tilde{\cal{R}}^{{\rm{DWBA}}}_{{\rm{post}}}}$] }\\
 \hline\\
  ${\rm {^9Be(^{10}B,^9Be)^{10}B}}$ &\,\,\,\,100 \cite{Mukh2}&\,\,\,\,\,\,\,\,\,\,\,\,\,\,\,0.339 - $i\cdot$2.664&\,\,\,\,\,\,\,-4.117&-0.193 - $i\cdot$1.521 \\
  &&\,\,\,\,\,\,\,\,\,\,\,(0.5154 - $i\cdot$4.0530)&&(-0.1270 - $i\cdot$0.9995)\\
   &&&&[1.521 + $i\cdot$1.300x$10^{-15}$]\\
 ${\rm {^{11}B(^{12}C,^{11}B)^{12}C}}$  
 \,\,\,\,\,\,\,&\,\,\,\, 87 \cite{LOLA}&\,\,\,\,\,\,\,\,\,\,\,\,\,-0.911 + $i\cdot$0.835&\,\,\,\,\,\,\,-1.714&\,\,1.023 + $i\cdot$0.937 \\
  &&\,\,\,\,\,\,\,\,\,\,\,(-1.260 + $i\cdot$1.154)&&(0.7399 + $i\cdot$0.6777)\\
    &&&&[1.382 + $i\cdot$5.200x$10^{-16}$]\\
   ${\rm{^{16}O(^3He}}, d{\rm {)^{17}F}}$(g.s.)&\,\,\, 29.75 \cite{Mukh6}&\,\,\,\,\,\,\,\,\,\,\,\,\,261.48 + $i\cdot$435.04&\,\,\,\,\,\,-590.36 &-0.599 + $i\cdot$0.996 \\
   &&\,\,\,\,\,\,\,\,\,\,\,(279.68 + $i\cdot$465.32)&&(-0560 + $i\cdot$0.932)\\
     &&&&[1.069 - $i\cdot$1.600x$10^{-15}$]\\
  ${\rm{^{16}O(^3He}}, d{\rm {)^{17}F}}$(0.429 MeV)& &\,\,\,\,\,\,\,\,\, (-2.96 - $i\cdot$4.75)x10$^{{\rm{15}}}$&\,\,\,\,\,-1.33x10$^{{\rm{9}}}$ &(1.26 -  $i\cdot$2.05)x10$^{{\rm{-7}}}$ \\
     &&\,\,\,\,\,\,\,\,\,\,\,((-0.725  - $i\cdot$1.160)x10$^{{\rm{15}}}$)&&((5.14  - $i\cdot$8.26)x10$^{{\rm{-7 }}}$)\\ 
    &&&&[0.245 - $i\cdot$1.300x$10^{-11}$]\\   
 ${\rm{^{19}F(}}p,\,\alpha {\rm {)^{16}O}}$&\,\,\,\, 0.250 \cite{HL1978}&\,\,\,\,\,\,\,(-1.360 + $i\cdot$0.453)x10$^{{\rm{-3}}}$&\,\,\,-1.68x10$^{{\rm{-3}}}$&\,\,1.112 + $i\cdot$0.370 \\
  &&\,\,\,\,\,\,\,\,\,\,\,((-0.148 + $i\cdot$4.940)x10$^{-4}$)&&(1.023 + $i\cdot$0.340)\\
    &&&&[1.088 - $i\cdot$7.150x$10^{-18}$]\\
  &\,\,\,\, 0.350&\,\,\,\,\,\,\,(-3.20 + $i\cdot$1.14 )x10$^{{\rm{-3}}}$&\,\,\,-3.98x10$^{{\rm{-3}}}$&\,\,1.104+ $i\cdot$0.394 \\
   &&\,\,\,\,\,\,\,\,\,\,\,((-3.480 - $i\cdot$1.240)x10$^{-3}$)&&(1.014 + $i\cdot$0.361)\\
     &&&&[1.088 - $i\cdot$6.005x$10^{-18}$]\\
 &\,\,\,\, 0.450&\,\,\,\,\,\,\,(-5.41 + $i\cdot$2.04 )x10$^{{\rm{-3}}}$&\,\,\,-6.78x10$^{{\rm{-3}}}$&\,\,1.097+ $i\cdot$0.412 \\ 
   &&\,\,\,\,\,\,\,\,\,\,\,((-5.893  - $i\cdot$2.217)x10$^{-3}$)&&(1.008 + $i\cdot$0.379)\\
    &&&&[1.088]\\ 
      &\,\,\,\,  0.327 \cite{Ivano2015}&\,\,\,\,\,\,\,(-2.750 + $i\cdot$0.97)x10$^{{\rm{-3}}}$&\,\,\,-3.42x10$^{{\rm{-3}}}$&\,\,1.110 + $i\cdot$0.390 \\
  &&\,\,\,\,\,\,\,\,\,\,\,((-3.00 + $i\cdot$1.05)x10$^{{\rm{-3}}}$)&&(1.020 + $i\cdot$0.360)\\
    &&&&[1.090]\\
     &\,\,\,\, 0.387&\,\,\,\,\,\,\,(-3.980 + $i\cdot$1.450 )x10$^{{\rm{-3}}}$&\,\,\,-4.97x10$^{{\rm{-3}}}$&\,\,1.100+ $i\cdot$0.400 \\
   &&\,\,\,\,\,\,\,\,\,\,\,((-4.330 - $i\cdot$1.580)x10$^{{\rm{-3}}}$)&&(1.010 + $i\cdot$0.370)\\
     &&&&[1.090]\\
      &\,\,\,\, 0.486&\,\,\,\,\,\,\,(-6.30 + $i\cdot$2.41 )x10$^{{\rm{-3}}}$&\,\,\,-7.90x10$^{{\rm{-3}}}$&\,\,1.090+ $i\cdot$0.420 \\ 
   &&\,\,\,\,\,\,\,\,\,\,\,((-6.850  + $i\cdot$2.62)x10$^{{\rm{-3}}}$)&&(1.010 + $i\cdot$0.380)\\
    &&&&[1.090]\\ 
    \hline\\
   \end{tabular}}
\end{center}
\end{table}}

\newpage

{\selectlanguage{english}
\begin{table}[t]
\begin{center}
 \caption{\label{table3}{\small  Reaction,  energy $E_x$,  set of the optical potentials (set), virtual decay $B\to\,A\,+\,a$,  orbital and total angular momentums ($l_B$, $j_B$),   square modulus of    the nuclear vertex constant $|G_B|^{{\rm {2}}}$( $G_B=G_{Aa;l_Bj_B}$) for the virtual decay   $B\to \,A\,+\,a$ and   the corresponding  ANC  $C_B^{{\rm {2}}}$ ($C_B=C_{Aa;l_Bj_B}$) $\,A\,+\,a\,\to\,B$. Figures in brackets are experimental and theoretical uncertainty, respectively, whereas those in square brackets are weighed mean derived from   the ANCs (NVCs) values  for  the sets 1 and 2. }}
 
 \vspace{2mm}
{\footnotesize
\begin{tabular}{lllllll}
 \hline
\multicolumn{1}{c}{$A$($x$, $y$)$B$}
 &\multicolumn{1}{c}{\,\,\,  $E_x$, MeV}
  &\multicolumn{1}{c}{ set}
 &\multicolumn{1}{c}{ $B\to\,A\,+\,a$}
&\multicolumn{1}{c}{\,\,$l_B$, $j_B$}
&\multicolumn{1}{c}{$|G_B|^{{\rm {2}}}$, fm}
&\multicolumn{1}{c}{$C_B^{{\rm {2}}}$, fm$^{{\rm{-1}}}$}\\
 \hline\\
 ${\rm {^9Be(^{10}B,^9Be)^{10}B}}$& 
    100 \cite{Mukh2}& 1& ${\rm{^{10}B}}\to\, {\rm{^9Be}} \,+p$&1, 3/2&  0.72($\pm$0.03;$\pm$0.02)&4.22($\pm$0.15;$\pm$0.10)\\
    &  & & & &0.72 $\pm$0.03&4.22$\pm$0.18\\
    &    & &  & &  0.84$\pm$0.03 \cite{Mukh2} &4.91$\pm$0.19 \cite{Mukh2}\\
&  & 2& & &0.77($\pm$0.03;$\pm$0.02)&4.49($\pm$0.16;$\pm$0.11)\\
  &  & & & &0.77 $\pm$0.03&4.49$\pm$0.19\\
   &  & &  & &  0.92$\pm$0.04 \cite{Mukh2} &5.35$\pm$0.21 \cite{Mukh2}\\
  &  & 1+2& & &[0.75($\pm$0.02;$\pm$0.02)]&[4.35($\pm$0.14;$\pm$0.14)]\\
    &  & & & &[0.75$\pm$0.03]&[4.35$\pm$0.19]\\
     &  & &  & &[0.87$\pm$0.08] \cite{Mukh2}&[5.06$\pm$0.46] \cite{Mukh2}\\
 ${\rm {^{11}B(^{12}C,^{11}B)^{12}C}}$ &87 \cite{LOLA} & &${\rm{^{12}C}}\to\, {\rm{^{11}B}} \,+p$ &1, 3/2&51.5($\pm$1.8;$\pm$1.3)&311.6($\pm$10.9;$\pm$7.7)\\
  &    & &  & &  51.5$\pm$2.2 &311.6$\pm$13.3\\
   ${\rm{^{16}O(^3He}}, d{\rm {)^{17}F}}$(g.s.) &29.75 \cite{Mukh6}& 1 &${\rm{^{17}F}}\to\, {\rm{^{16}O}} \,+p$ &2, 5/2&0.179($\pm$0.018;$\pm$0.009)&1.14($\pm$0.12;$\pm$0.06)\\
    &    & &  & &  0.179$\pm$0.020   &1.14$\pm$0.13  \\
    &    & &  & &  0.16 \cite{Mukh6} &1.0 \cite{Mukh6}\\
     & & 2 & & &0.206($\pm$0.021;$\pm$0.010)&1.31($\pm$0.14;$\pm$0.07)\\
      &    & &  & &  0.206$\pm$0.024   &1.31$\pm$0.15  \\
       &    & &  & &  0.18 \cite{Mukh6}   &1.10  \cite{Mukh6} \\
 &  &1+2 & & &[0.190($\pm$0.014;$\pm$0.013)]&[1.21($\pm$0.09;$\pm$0.08)]\\
     &    & &  & & [0.190$\pm$0.019]   &[1.21$\pm$0.12]  \\
        &    & &  & &  [0.170$\pm$0.016] \cite{Mukh6}   &[1.08$\pm$0.10] \cite{Mukh6} \\
              &18;34 \cite{Ar96}     & &  & & 0.16  \cite{Ar96}   &1.02 \cite{Ar96}\\
 ${\rm{^{16}O(}}p, \gamma{\rm {)^{17}F}}$(g.s) &     & &  & &  0.17$\pm$0.02 \cite{Ar2009}   &1.09$\pm$0.11  \cite{Ar2009} \\   
     ${\rm{^{16}O(^3He}}, d{\rm {)^{17}F}}^*$  &29.75 \cite{Mukh6}& 1 &${\rm{^{17}F}}\to\, {\rm{^{16}O}} \,+p$ &0, 1/2&916($\pm$96;$\pm$46)&5840($\pm$611;$\pm$292)\\
     &    & &  & &  916$\pm$106  &5840$\pm$667 \\
      &    & &  & &  939 \cite{Mukh6} &5980 \cite{Mukh6}\\
     & & 2 & & &1053($\pm$110;$\pm$53)&6713($\pm$703;$\pm$335)\\
      &    & &  & & 1053$\pm$122&6713$\pm$779\\
           &    & &  & & 1099 \cite{Mukh6} &7000 \cite{Mukh6}\\
            &    &1+2 &  & & [975($\pm$72;$\pm$68)]&[6216($\pm$461;$\pm$432)]\\
                      &    &  &  & & [975$\pm$99]   &[6216$\pm$632]  \\
        &    & &  & &  [1019$\pm$107] \cite{Mukh6}   &[6490$\pm$680] \cite{Mukh6} \\
                    &18; 34  \cite{Ar96}    & &  & & 840; 819 \cite{Ar96}   &5355; 5122 \cite{Ar96}\\
 ${\rm{^{16}O(}}p, \gamma{\rm {)^{17}F}}^*$ &     & &  & &  893$\pm$35 \cite{Ar2009}   &5700$\pm$225  \cite{Ar2009} \\            
     ${\rm{^{19}F(}}p,\,\alpha {\rm {)^{16}O}}$  &0.250  \cite{HL1978}&  &${\rm{^{19}F}}\to\, {\rm{^{16}O}} \,+\,t$ &0, 1/2&13.5($\pm$2.1;$\pm$0.7)&618.1($\pm$95.2;$\pm$30.9)\\
   &0.350   &  & & &13.2($\pm$1.4;$\pm$0.7)&605.0($\pm$63.4;$\pm$30.3)\\    
 &0.450   &  & & &11.9($\pm$1.3;$\pm$0.6)&544.8($\pm$60.6;$\pm$27.2)\\
  weighed mean &
  &  & &  &12.7($\pm$0.9;$\pm$0.5)&583.5($\pm$39.8;$\pm$23.3)\\                
  & &  &  &  &12.7$\pm$1.0&583.5$\pm$46.1\\
  &0.327  \cite{Ivano2015}&  &  & &28.1($\pm$2.7;$\pm$1.4)&1290.3($\pm$124.0;$\pm$64.3)\\
   &0.387   &  & & &29.2($\pm$3.2;$\pm$1.5)&1341.3($\pm$144.7;$\pm$66.6)\\    
 &0.486   &  & & &27.2($\pm$2.9;$\pm$1.4)&1248.1($\pm$134.6;$\pm$62.5)\\
 weighed mean &
  &  & &  &28.1($\pm$1.7;$\pm$0.8)&1291.1($\pm$77.2;$\pm$37.2)\\                
  & &  &  &  &28.1$\pm$1.9&1291.1$\pm$85.7\\                
  \hline
 \end{tabular}}
\end{center}
\end{table}

 \end{document}